\documentclass{article} 
\usepackage{iclr2027_conference,times}

\usepackage{amsmath,amsfonts,bm}

\def\eqref#1{equation~\ref{#1}}

\def\1{\bm{1}}

\DeclareMathAlphabet{\mathsfit}{\encodingdefault}{\sfdefault}{m}{sl}
\SetMathAlphabet{\mathsfit}{bold}{\encodingdefault}{\sfdefault}{bx}{n}

\usepackage{graphicx}
\usepackage{hyperref}
\usepackage{url}
\usepackage{enumitem}
\usepackage{wrapfig, subfigure, subcaption}
\usepackage{microtype}

\usepackage{changepage}

\usepackage{xcolor}
\usepackage{soul}
\usepackage[normalem]{ulem}

\makeatletter
\font\uwaveboldfont=lasyb10 scaled 750
\newcommand{\bolduwave}[1]{%
  \bgroup
  \markoverwith{%
    \lower3.5\p@\hbox{\uwaveboldfont\char58}%
  }%
  \ULon{#1}%
}
\makeatother

\definecolor{contextunderline}{RGB}{0,0,0}
\newcommand{\context}[1]{%
  \textcolor{contextunderline}{\textbf{\bolduwave{#1}}}%
}

\newcommand{\operation}[1]{%
  {\setul{0.5ex}{0.6pt}%
  \textcolor{black}{\textbf{\ul{#1}}}}%
}

\usepackage[scaled=0.92]{inconsolata}

\usepackage{caption}
\hypersetup{hidelinks}

\usepackage{booktabs}
\usepackage{multirow}
\usepackage{makecell}

\usepackage[table]{xcolor}
\definecolor{ContextBlue}{RGB}{225,241,250}
\newcommand{\bluecell}[1]{\cellcolor{ContextBlue}#1}
\definecolor{OperationGreen}{RGB}{230,241,231}
\newcommand{\greencell}[1]{\cellcolor{OperationGreen}#1}

\usepackage[most]{tcolorbox}
\usepackage{xcolor}
\definecolor{myblue}{RGB}{62,119,213}
\definecolor{myorange}{RGB}{238,135,66}

\usepackage{eso-pic}

\AddToShipoutPictureFG*{%
  \AtPageLowerLeft{%
    \hspace*{\dimexpr 1in+\oddsidemargin\relax}%
    \raisebox{0.45in}{%
    \normalfont\small\itshape (Preprint. Under review)%
    }%
  }%
}

\title{Ask Without Telling: Local SLMs Consult \\ Cloud LLMs Without Revealing Task Intent}

\iclrfinalcopy

\author{Yanmeng Wang$^1$, Yunxuan Li$^2$, Shilong Fan$^1$, Yuhan Zheng$^1$, Tsung-Hui Chang$^{2*}$\\
$^1$Nanjing University of Posts and Telecommunications, Nanjing, 210023, China\\
$^2$The Chinese University of Hong Kong, Shenzhen, 518172, China\\
$^*$Corresponding author.
}

\begin{document}

\maketitle

\fancyhead{}
\renewcommand{\headrulewidth}{0.4pt}

\vspace{-23pt}
\begin{abstract}
\vspace{-3pt}
\begin{adjustwidth}{-0.24in}{-0.24in}
As local small language models (SLMs) increasingly collaborate with more capable cloud large language models (LLMs), a natural privacy question arises: \emph{Can a local SLM obtain cloud LLM guidance while protecting user privacy?}
Existing privacy-preserving SLM-LLM frameworks primarily hide \emph{sensitive values} while preserving task semantics, which can still expose what the user is trying to accomplish.
For example, allocating scarce medical supplies across hospitals may signal an emerging public-health emergency, while rebalancing an investment portfolio may reveal a private investment strategy, even when names and numerical values are hidden.
Recent decoy-based methods further obscure task intent by hiding the real request among alternatives, but stronger protection relies on more decoys or semantic abstraction, increasing overhead or risking utility loss.
More fundamentally, existing work does not systematically characterize the components of private task intent or how each should be protected.
We therefore introduce \emph{task-private consultation}, which characterizes task intent through two components: \emph{task context} and \emph{task operation}.
To the best of our knowledge, this is the first systematic study of these components and their individual and joint protection in local-cloud SLM-LLM consultation.
To realize this setting, we propose \texttt{PriCon}, an end-to-end framework that transforms the task itself through recoverable mathematical reformulation rather than hiding it among alternatives.
A local closed-loop refinement mechanism further maintains privacy and recoverability throughout consultation.
Experiments on 100 tasks show that \texttt{PriCon} reduces cloud-side task-intent inference Hit@1 to nearly 0\%, versus 93--99\% under sensitive-value removal and 3--30\% under decoy-based protection, while preserving cloud-assisted utility.
\end{adjustwidth}
\end{abstract}

\vspace{-10pt}
\section{Introduction} \label{sec:introduction}

\vspace{-4pt}
Imagine a local small language model (SLM) assistant deployed by a regional public-health authority.
Following a sudden increase in infectious-disease cases, the authority must allocate a limited stock of critical medical supplies across hospitals.
Different hospitals face different supply demands and urgency levels, while the total regional stock is limited.
The local SLM may not have sufficient reasoning capability to reliably coordinate the allocation across all hospitals, and therefore turns to a more capable cloud-hosted large language model (LLM) for assistance.

Before consulting the cloud, the SLM removes or replaces sensitive values, including hospital names, priority weights, inventory levels, and demand values.
The new request appears sanitized:
\begin{tcolorbox}[
    breakable,
    colback=black!2,
    colframe=black!45,
    boxrule=0.7pt,
    arc=3mm,
    left=4pt,
    right=4pt,
    top=3pt,
    bottom=3pt
]

\small\ttfamily
\linespread{1.05}\selectfont
\textls[-15]{
``%
Given \context{hospitals} [H] with different priority levels [W], multiple types of \context{medical supplies} [S] with limited available stocks [B], and hospital-specific \context{supply demands} [D], 
\operation{allocate the available supplies} across hospitals to minimize the overall \operation{impact of medical-supply shortages}.%
''
}
\end{tcolorbox}
No hospital name, exact demand, inventory level, or priority value is revealed.
Yet the request still tells a revealing story.
It exposes both the \emph{context}: a public-health authority distributing scarce medical resources across hospitals, and the \emph{operation}: a constrained allocation problem prioritizing hospitals while minimizing shortage impact.
From this alone, an observer may infer unusual pressure on regional medical resources, potentially signaling an emerging public-health emergency.

Such task-intent leakage extends well beyond public-health settings.
For example, rescheduling a production line may reveal an upcoming product launch or a capacity bottleneck; 
restructuring a small business's debt may signal severe cash-flow pressure; 
and rebalancing an investment portfolio may expose a confidential investment strategy. 
In each case, hiding names, identifiers, and numerical values is not enough. 
\emph{What the user aims to accomplish may itself be private}.

This risk is particularly relevant to the emerging paradigm of local SLM-cloud LLM collaboration, where local models work together with more capable cloud models to solve tasks that cannot be reliably completed on-device alone.
Recent systems therefore split responsibilities between local SLMs that handle private data and local execution, and stronger cloud LLMs that provide reasoning and planning guidance for complex tasks~\citep{zhang-etal-2024-cogenesis,narayan2025minions,ni2026following}. 
However, obtaining such guidance still requires the local SLM to send task-related information to the cloud, making every consultation message visible to the cloud provider.

Existing privacy-aware collaboration primarily protects \emph{sensitive values} by removing, replacing, or obscuring private information before cloud consultation~\citep{siyan-etal-2025-papillon,yuan2026paac,zhan2026prism}.
However, as the medical-supply example illustrates, preserving task semantics for cloud assistance can still expose the application context, requested operation, and underlying objective.
Recent decoy-based methods further obscure user intent by hiding the true request among plausible alternatives~\citep{huang2026homellama,zhou2026gs}.
Yet the true request remains cloud-visible, while stronger obfuscation requires more decoys or semantic abstraction, increasing inference overhead or potentially compromising task utility.
More fundamentally, existing approaches do not systematically characterize the components of private task intent or how each should be protected.

This gap motivates a fundamental question:

\vspace{-4pt}
\begin{quote} 
\setlength{\leftskip}{-0.7em}
\setlength{\rightskip}{-0.7em}
\emph{Can a local SLM consult a cloud LLM without revealing the user's private task intent?}
\end{quote}
\vspace{-4pt}

We take a more structured view and refer to this setting as \textbf{task-private consultation}.
We view task intent as containing two components: \emph{task context}, which captures the real-world setting from which the task arises, and \emph{task operation}, which captures the computation or decision the user wishes to perform.
Either component may be private.
Importantly, the two are not independent sources of leakage: contextual details can themselves reveal the underlying operation.

\begin{wrapfigure}{r}{0.6\textwidth}  
\centering
\vspace{-15pt}
\includegraphics[width=0.6\textwidth]{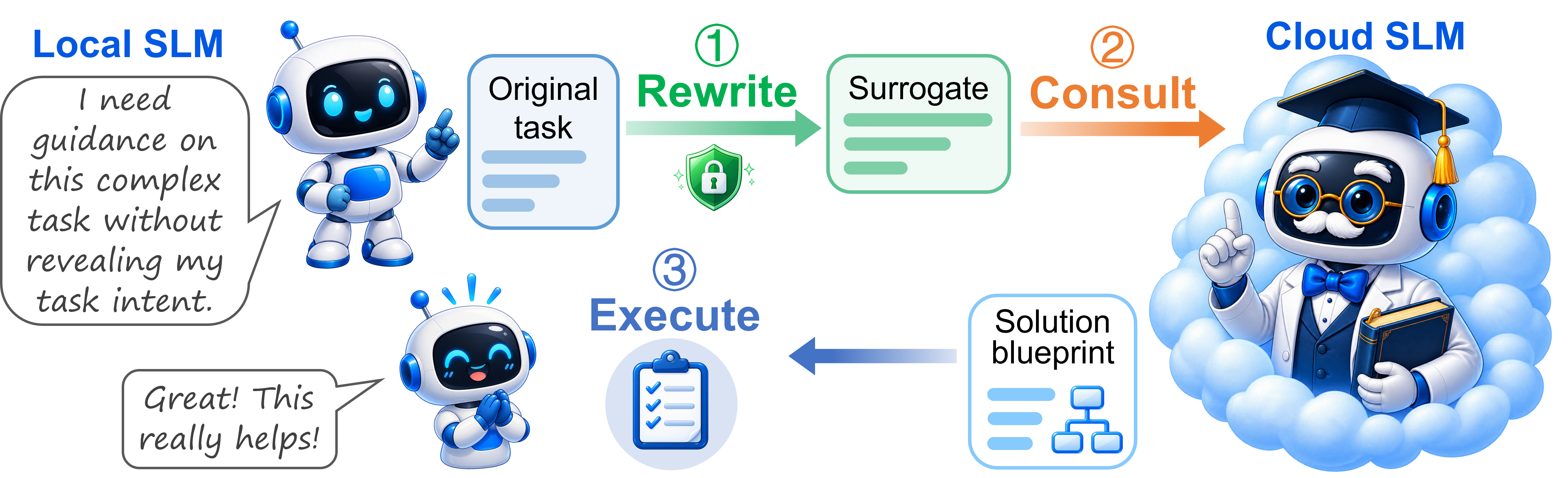}
\vspace{-18pt}
\caption{Task-private consultation.}
\vspace{-10pt}
\label{fig:task-private-consultation}
\end{wrapfigure}
As illustrated in Fig.~\ref{fig:task-private-consultation}, task-private consultation follows a
\emph{rewrite, consult, and execute} workflow:
\emph{1) Rewrite}: the local SLM transforms the original task into an intent-obscuring surrogate;
\emph{2) Consult}: the cloud LLM reasons only over the surrogate and returns a transferable solution blueprint;
\emph{3) Execute}: the local SLM maps the blueprint back to the original task and executes the resulting solution locally.
If further assistance is needed, local difficulties are converted into surrogate-level diagnostic requests and the consultation is repeated without exposing the original task or local execution details.
We formalize this workflow, its protection scopes, and the cloud-visible transcript in Section~\ref{sec:task-private-consultation}.

Realizing this workflow presents two central challenges:

\vspace{-4pt}
\begin{itemize}[leftmargin=1.5em, itemsep=0pt, topsep=3pt]
\item
\emph{How can the local SLM construct an intent-obscuring surrogate?}
The surrogate must hide the protected intent while preserving the solution-relevant structure needed for cloud guidance.
Insufficient transformation leaves the intent recognizable, whereas excessive transformation may weaken the structure needed for useful guidance.

\item
\emph{How can privacy and recoverability be maintained throughout the consultation?}
A safe initial rewrite is insufficient, as local execution failures may trigger follow-up requests that reveal additional task-specific information.
Each consultation round must therefore preserve both privacy and local recoverability. 

\end{itemize}
\vspace{-4pt}

To address these challenges, we propose \texttt{PriCon}, an end-to-end framework for task-private consultation.
Instead of hiding the true task among alternatives, \texttt{PriCon} transforms the task itself through recoverable mathematical reformulation.
For context protection, it extracts a domain-neutral mathematical representation that preserves solution-relevant structure and then uses semantic rewriting to recast it as a different application story.
For operation or joint protection, it additionally changes the apparent computation through an alternative mathematical formulation before semantic rewriting.
A local review and \emph{closed-loop refinement} mechanism further preserves privacy and recoverability throughout the consultation while reusing successful experience across related tasks.

In summary, our main contributions are as follows:

\vspace{-3pt}
\begin{itemize}[leftmargin=1.5em, itemsep=0pt, topsep=2pt]

\item \textbf{Task-private consultation.}
We systematically characterize task intent through \emph{task context} and \emph{task operation}, reveal their coupled leakage, and define context, operation, and joint protection over the complete cloud-visible transcript.

\item \textbf{PriCon framework.}
We propose \texttt{PriCon}, which transforms the original task into a recoverable, intent-obscuring surrogate through domain-neutral mathematical reformulation and semantic rewriting, while keeping the recovery mapping local. 
To the best of our knowledge, \texttt{PriCon} is the first local-cloud LLM consultation framework to use recoverable mathematical reformulation for task-intent protection without relying on additional decoy queries.

\item \textbf{Transcript-level intent evaluation.}
We evaluate task-intent inference by an honest-but-curious cloud LLM from the complete cloud-visible transcript.
Experiments on 100 tasks show that \texttt{PriCon} strongly protects task intent while preserving the utility of cloud LLM assistance.

\end{itemize}

\vspace{-3pt}
\section{Related Work} \label{sec:related_work}

\vspace{-3pt}
We organize prior work around the evolving goals of local-cloud language-model collaboration, from efficient workload sharing to privacy-aware delegation and task-intent protection.

\textbf{Local-cloud language-model collaboration.}
Existing local SLM-cloud LLM systems divide work according to query difficulty, reasoning complexity, data locality, or system cost.
\texttt{Hybrid LLM}~\citep{ding2024hybridllm} routes queries between local SLMs and cloud LLMs according to predicted difficulty and desired quality, while \texttt{AdaSwitch}~\citep{sun-etal-2024-adaswitch} lets a local SLM handle simpler reasoning steps and escalates harder ones to the cloud LLM.
\texttt{Minions}~\citep{narayan2025minions} keeps long contexts local while using the cloud LLM to decompose tasks into simpler subtasks for local execution.
\texttt{CE-CoLLM}~\citep{jin2025cecollm} and \texttt{TMO}~\citep{yuan2026tmo} further optimize local-cloud execution for latency, communication, and inference cost.
Overall, these approaches primarily treat local-cloud collaboration as a capability- and efficiency-allocation problem.
As such collaboration involves user-side context and intermediate task information, privacy becomes an important concern.

\textbf{Privacy-preserving SLM-LLM collaboration.}
Recent privacy-aware frameworks use local SLMs to limit private information exposed to the cloud, either by keeping private context local or sanitizing transmitted information.
In its sketch-based variant, \texttt{CoGenesis}~\citep{zhang-etal-2024-cogenesis} keeps personalized context local and sends only a general task instruction to the cloud LLM, which returns a sketch for local generation.
\texttt{Need to Know}~\citep{huang2026needtoknow} suppresses task-unnecessary private information while preserving task-essential information for cloud reasoning.
Other approaches sanitize private information before transmission.
\texttt{PAPILLON}~\citep{siyan-etal-2025-papillon} rewrites private queries into privacy-preserving prompts, \texttt{PAAC}~\citep{yuan2026paac} replaces sensitive values with typed placeholders, and \texttt{PRISM}~\citep{zhan2026prism} perturbs sensitive entities using local differential privacy.
Despite their different mechanisms, these approaches primarily reduce the exposure of private information while preserving the task semantics needed for cloud assistance.
Such semantic preservation can still leave the task context and requested operation recognizable, allowing the cloud to infer private activities or objectives even when sensitive values are protected.

\textbf{Task-intent protection.}
Recent work has begun to protect task intent through decoy-based obfuscation.
\texttt{PrivShield}~\citep{huang2026homellama} hides a real query among adversarial queries, while \texttt{GS-Chaff}~\citep{zhou2026gs} combines query-adaptive semantic abstraction with plausible decoy queries.
Despite their different designs, both rely on ambiguity, with a query representing the real request remaining in a cloud-visible candidate set alongside decoys.
Achieving stronger obfuscation may require more decoys or semantic abstraction, increasing inference overhead or potentially compromising task utility.
In contrast, \texttt{PriCon} replaces the original task with a recoverable surrogate through mathematical reformulation, without sending the original task or decoys to the cloud.

\vspace{-3pt}
\section{Task-Private Consultation} 
\label{sec:task-private-consultation}

\vspace{-3pt}
We now formalize task-private consultation by specifying the consultation setting, the information to be protected, and the threat model considered in this work.

\vspace{-2pt}
\subsection{Consultation Setting and Workflow} \label{sec:consultation-setting}

\vspace{-2pt}
We consider a trusted local device hosting an SLM and a more capable cloud-hosted LLM.
The local SLM receives an original task $\mathcal{T}$, sensitive values $\mathcal{V}$, and access to a local execution environment.

Here, $\mathcal{V}$ denotes task-specific sensitive inputs, such as names, exact locations, identifiers, and measurements, that are required to instantiate the task.

We focus on tasks that exceed the SLM's end-to-end reasoning capability but can be reformulated into a recoverable surrogate whose solution can be instantiated locally.
For such tasks, the SLM consults the cloud without transmitting the original task.
As illustrated in Fig.~\ref{fig:task-private-consultation}, task-private consultation follows a \emph{rewrite, consult, and execute} workflow:
\vspace{3pt}
$$
\underbrace{\mathcal{T}}_{\text{\fontsize{7.5pt}{9pt}\selectfont Original Task}}
\!\!\!\!
\xrightarrow[\mathcal{M}]{\text{\fontsize{7.5pt}{9pt}\selectfont (1) SLM: Rewrite}}
\!\!\!\!\!
\underbrace{\widetilde{\mathcal{T}}}_{\text{\fontsize{7.5pt}{9pt}\selectfont Intent-Obscuring Surrogate}}
\!\!\!\!\!
\xrightarrow{\text{\fontsize{7.5pt}{9pt}\selectfont (2) LLM: Consult}}
\!\!\!\!\!
\underbrace{\mathcal{B}}_{\text{\fontsize{7.5pt}{9pt}\selectfont Solution Blueprint}}
\!\!\!\!
\xrightarrow[\mathcal{M}, \mathcal{V}]{\text{\fontsize{7.5pt}{9pt}\selectfont (3) SLM: Execute}}
\!\!\!\!
\underbrace{\widehat{\mathcal{Y}}}_{\text{\fontsize{7.5pt}{9pt}\selectfont Local Solution}}
\!\!
.
$$
\vspace{-8pt}

\begin{enumerate}[label=\arabic*), leftmargin=1.5em, itemsep=0.5pt, topsep=2pt]

\item \textbf{Rewrite.}
The local SLM transforms the original task $\mathcal{T}$ into an intent-obscuring surrogate $\widetilde{\mathcal{T}}$ while replacing sensitive values $\mathcal{V}$ with placeholders.
It also establishes a local correspondence $\mathcal{M}$ recording how solution-relevant elements of the surrogate relate to the original task.

\item \textbf{Consult.}
The cloud LLM receives only $\widetilde{\mathcal{T}}$ and returns a transferable solution blueprint $\mathcal{B}$, such as an algorithm, decision procedure, program skeleton, or reasoning steps.
The blueprint is expressed entirely in terms of the surrogate because the cloud does not observe $\mathcal{T}$, 
$\mathcal{M}$, or $\mathcal{V}$.

\item \textbf{Execute.}
The local SLM uses the correspondence $\mathcal{M}$ to map the cloud-provided blueprint $\mathcal{B}$ back to the original task and instantiates it with the sensitive values $\mathcal{V}$. 
The resulting solution $\widehat{\mathcal{Y}}$ is verified against the original task requirements and executed locally.
If further assistance is needed during local interpretation or execution, the SLM constructs a surrogate-level diagnostic request and repeats the consultation without exposing the original task or execution trace.

\end{enumerate}

\vspace{-2pt}
A consultation may contain an initial request followed by multiple diagnostic rounds.
We define the \emph{cloud-visible transcript} as the complete sequence of surrogate requests and cloud responses exchanged throughout the interaction.
Crucially, task-intent privacy is defined over this complete transcript rather than over any single sanitized request.
The original task, sensitive values, correspondence, and task-specific execution traces are excluded from this transcript and remain on the local device.

\subsection{Task Intent and Protection Scope} 
\label{sec:task-intent} 

Beyond protecting sensitive values $\mathcal{V}$, task-private consultation aims to protect the task intent revealed through cloud interaction.
We use the medical-supply allocation example from Section~\ref{sec:introduction} as a running example to formalize this notion.
We characterize task intent through two components:

\vspace{-3pt}
\begin{enumerate}[label=\arabic*), leftmargin=1.5em, itemsep=0.5pt, topsep=2pt]
\item
\textbf{Task context} $\mathcal{C}$ captures the application domain, entities, and situational background.
In our example, it corresponds to a public-health setting involving hospitals, medical-supply demands, and scarce regional supplies.

\item 
\textbf{Task operation} $\mathcal{O}$ captures the computation or decision requested by the user, including its objective and major constraints.
Here, it corresponds to allocating limited medical supplies across hospitals to minimize priority-weighted shortage impact.
\end{enumerate}

\vspace{-2pt}
The user or application policy may designate task context $\mathcal{C}$, task operation $\mathcal{O}$, or both as the privacy target.
Accordingly, we consider three protection scopes:

\vspace{-3pt}
\begin{itemize}[leftmargin=1.5em, itemsep=0pt, topsep=2pt]
\item 
\textbf{Context protection} protects task context $\mathcal{C}$ without requiring task operation $\mathcal{O}$ itself to be hidden.
In the public-health scenario, the authority may wish to conceal the underlying medical emergency while still allowing the cloud to recognize that it is solving a generic resource-allocation problem.

\item
\textbf{Operation protection} protects task operation $\mathcal{O}$, while task context $\mathcal{C}$ need not itself be private.
For the medical-supply allocation example, if the public-health situation is already known, the privacy target may instead be how scarce supplies are prioritized and allocated across hospitals.

\item 
\textbf{Joint protection} protects both $\mathcal{C}$ and $\mathcal{O}$.
This applies when revealing either the underlying healthcare situation or the resource-allocation strategy would disclose private information.
\end{itemize}

\subsection{Threat Model}
\label{sec:threat-model}

\textbf{Trusted local device.}
We assume that the local SLM and all local components, including storage, rewriting, review, and execution procedures, behave as intended.
The original task $\mathcal{T}$, sensitive values $\mathcal{V}$, local correspondence $\mathcal{M}$, and local execution results remain under local control.
Compromise of the local device is outside the scope of this work.

\textbf{Honest-but-curious cloud.}
The cloud follows the consultation protocol and provides guidance, but may store and jointly analyze the complete cloud-visible transcript to infer the protected task intent.
Under this threat model, task-private consultation aims to make the protected task intent difficult to infer from the cloud-visible transcript.

\vspace{-2pt}
\section{PriCon: Realizing Task-Private Consultation}
\label{sec:Rewrite}
\vspace{-2pt}

We now present \texttt{PriCon}, our end-to-end framework for realizing task-private consultation.
It centers on two local mechanisms: \emph{intent-obscuring rewriting}, which constructs the cloud-visible surrogate, and \emph{closed-loop refinement}, which maintains privacy and recoverability throughout consultation.

\subsection{Intent-Obscuring Rewriting}
\label{sec:intent-rewrite}

\begin{wrapfigure}{r}{0.25\textwidth}  
\centering
\vspace{-12pt}
\includegraphics[width=0.25\textwidth]{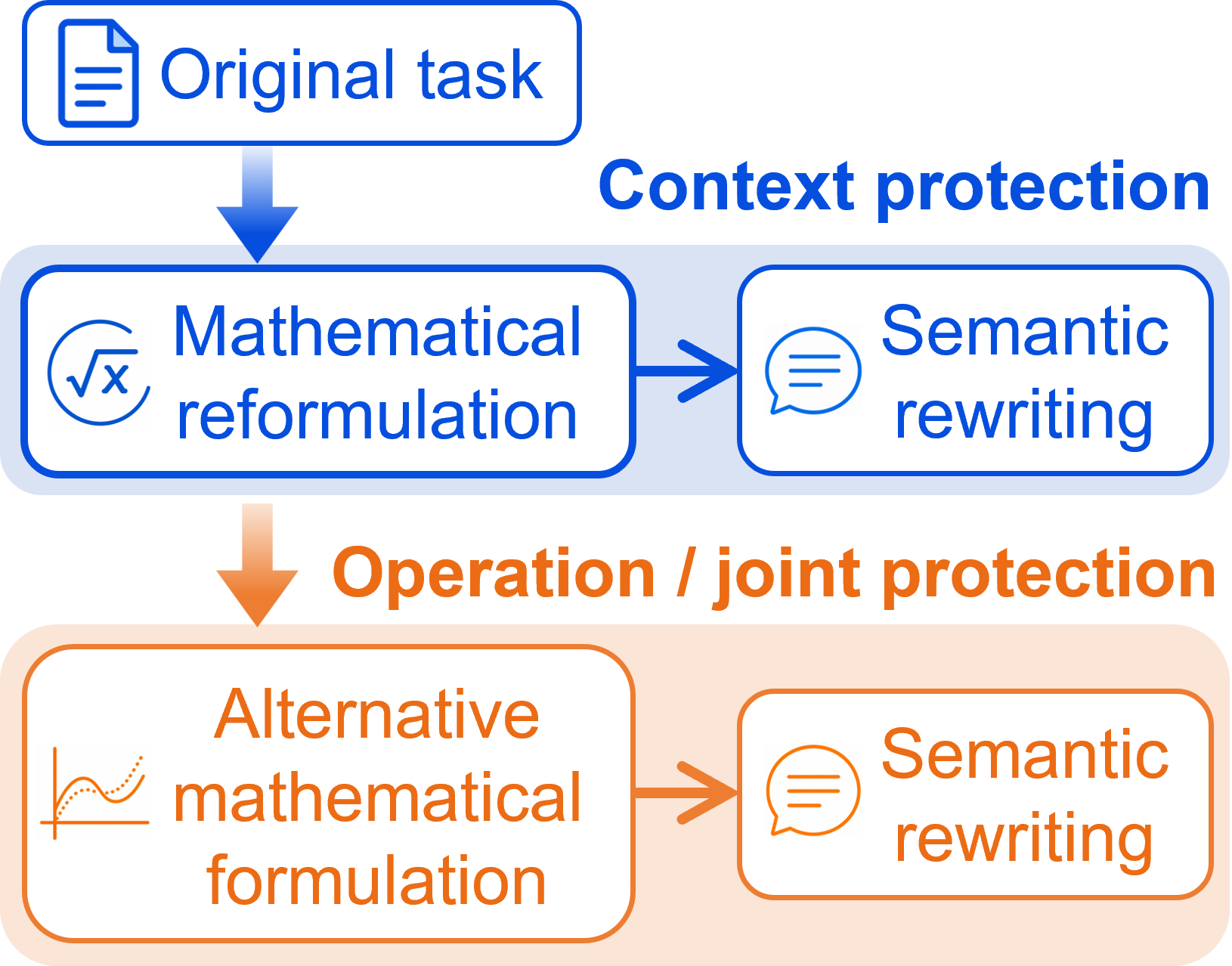}
\vspace{-17pt}
\caption{Rewriting paths.}
\label{fig:Rewrite}
\vspace{-10pt}
\end{wrapfigure}

As illustrated in Fig.~\ref{fig:Rewrite}, \texttt{PriCon} selects a rewriting path according to the protection scope defined in Section~\ref{sec:task-intent}.
A natural first step is to abstract the task into a domain-neutral mathematical representation.
Although this removes much of the original semantics, the resulting representation may still contain recognizable structural cues about the original task intent, as reflected by the \emph{Math-only} results in Table~\ref{tab:main-results}.
Rather than bypassing this step, \texttt{PriCon} retains the mathematical representation as a local intermediate, since it provides the \emph{experience bank} in Section~\ref{sec:privacy-review} with a shared structure for identifying similar tasks across domains and reusing successful experience.
It then applies semantic rewriting to recast this representation as a different application story before cloud consultation.
For context protection, \texttt{PriCon} follows this mathematical-to-semantic path; for operation or joint protection, it additionally introduces an \emph{alternative mathematical formulation} to change the apparent computation before semantic rewriting.
Throughout the process, the correspondence $\mathcal{M}$ remains local and maps cloud guidance back to the original task.
We illustrate these rewriting paths using our running medical-supply example.

\subsubsection{Context Protection}

We first consider context protection, where the goal is to hide the original task context while leaving the underlying task operation $\mathcal{O}$ unchanged.

\textbf{Mathematical reformulation.}
\texttt{PriCon} first expresses the task in a domain-neutral mathematical form, replacing domain-specific entities with abstract roles while preserving their relations and underlying operation.
Applying this abstraction to our running medical-supply example yields the following intermediate representation:

\begin{tcolorbox}[
    breakable,
    colback=myblue!5,
    colframe=myblue,
    boxrule=0.7pt,
    arc=3mm,
    left=4pt,
    right=4pt,
    top=3pt,
    bottom=3pt
]

\setlength{\leftskip}{0.1em}
\setlength{\rightskip}{0.1em}
\small\ttfamily
\linespread{1.05}\selectfont
``%
Given \context{demand groups} [G] with different priority weights [W], multiple \context{resource types} [R] with limited capacities [B], and group-specific resource requirements [D],
\operation{allocate available resources} across demand groups to minimize the overall \operation{impact of unmet resource requirements}.%
''
\end{tcolorbox}

Hospitals are abstracted as demand groups and medical supplies as resource types, while sensitive values remain local in $\mathcal{V}$.
This abstraction removes most healthcare-specific cues and substantially reduces task-intent inference, but does not eliminate it.
The remaining concepts, such as demand groups, limited capacities, and unmet requirements, still reveal a recognizable resource-allocation operation, leaving the cloud LLM with a few clues about the task context.
\texttt{PriCon} therefore treats the mathematical form as an intermediate representation rather than the final cloud-facing request.

\textbf{Semantic rewriting.}
\texttt{PriCon} further embeds the mathematical structure in a coherent but semantically distant application story.
Intuitively, this alternative story gives the cloud LLM a coherent surrogate-domain interpretation, making it less likely to associate the request with original task.
For example, the generic resource-allocation problem can be rewritten as GPU allocation in a data-center setting:
\begin{tcolorbox}[
    breakable,
    colback=myblue!5,
    colframe=myblue,
    boxrule=0.7pt,
    arc=3mm,
    left=4pt,
    right=4pt,
    top=3pt,
    bottom=3pt
]

\setlength{\leftskip}{0.5em}
\setlength{\rightskip}{0.5em}
\small\ttfamily
\linespread{1.05}\selectfont
``%
Given \context{model-training jobs} [J] with different priority levels [W], multiple types of \context{GPU resources} [G] with limited available capacities [B],
and job-specific compute demands [D],
\operation{allocate the available GPU resources} across training jobs to minimize the overall \operation{impact of unmet compute demand}.%
''
\end{tcolorbox}

Here, demand groups and resource types become model-training jobs and GPU resources.
The cloud now sees a GPU-allocation task rather than the original healthcare task, while the relationships needed for local recovery remain available through the correspondence $\mathcal{M}$.

\subsubsection{Operation Protection and Joint Protection}

Protecting task operation is more challenging because the surrounding context may itself reveal the protected operation.
For example, hospitals, medical-supply demands, and limited stocks already suggest a resource-allocation problem even if the operation is not explicitly stated.
\texttt{PriCon} therefore adopts a conservative strategy: whenever task operation $\mathcal{O}$ is protected, it also transforms contextual cues that could help infer $\mathcal{O}$.
Operation protection and joint protection consequently follow the same rewriting path, although their requested protection scopes remain conceptually distinct.

\textbf{Alternative mathematical formulation.}
Starting from the mathematical representation, \texttt{PriCon} constructs an alternative formulation with a different apparent computation and a recoverable solution mapping.
In our example, the generic resource-allocation problem is recast as shortest-path search on layered directed acyclic graphs (DAGs):
\begin{tcolorbox}[
    breakable,
    colback=myorange!5,
    colframe=myorange,
    boxrule=0.7pt,
    arc=3mm,
    left=4pt,
    right=4pt,
    top=3pt,
    bottom=3pt
]

\small\ttfamily
\linespread{1.1}\selectfont
\textls[-25]{
``%
Given a set of \context{layered directed graphs} [G], each with \context{states} [V], \context{feasible transitions} [E], and \context{transition costs} [C],
find a \operation{minimum-cost path} from the source to a terminal node in each graph.%
''
}
\end{tcolorbox}

The correspondence between the original and alternative formulations can be made explicit.
Since supply types are independent in both the objective and stock constraints, the task can be decomposed into separate allocation subproblems, one for each supply type.
For supply type $s$, the subproblem is:

\vspace{-8pt}
\begin{small}
\begin{equation}
\underbrace{
\min_{\{x_{h}^{s}\}}\;
{\sum}_{h=1}^{H}
w_h \cdot \ell_{h}^{s} (d_{h}^{s}-x_{h}^{s})
}_{\text{\fontsize{8pt}{8pt}\selectfont minimize shortage impact}} \, ,
\quad\;
\text{s.t.}
\underbrace{0\leq x_{h}^{s}\leq d_{h}^{s}, \;
\ x_{h}^{s} \in \mathbb{Z}_{\geq 0},\; \forall h}_{\text{\fontsize{8pt}{8pt}\selectfont allocation within demand, discrete units}}
\!\!\!;
\quad
\underbrace{{\sum}_{h=1}^{H}x_h^{s}\leq B^{s}}_{\text{\fontsize{8pt}{8pt}\selectfont stock constraint}}.
\label{eq:allocation}
\end{equation}
\end{small}
\vspace{-7pt}

\noindent
Here, $d_h^s$ and $x_h^s$ denote the demand and allocation of supply $s$ at hospital $h$, $w_h$ is its priority weight, $B^s$ is the available stock, and the function $\ell_h^s$ measures shortage impact.

\begin{wrapfigure}{r}{0.45\textwidth}  
\centering
\vspace{-11pt}
\includegraphics[width=0.45\textwidth]{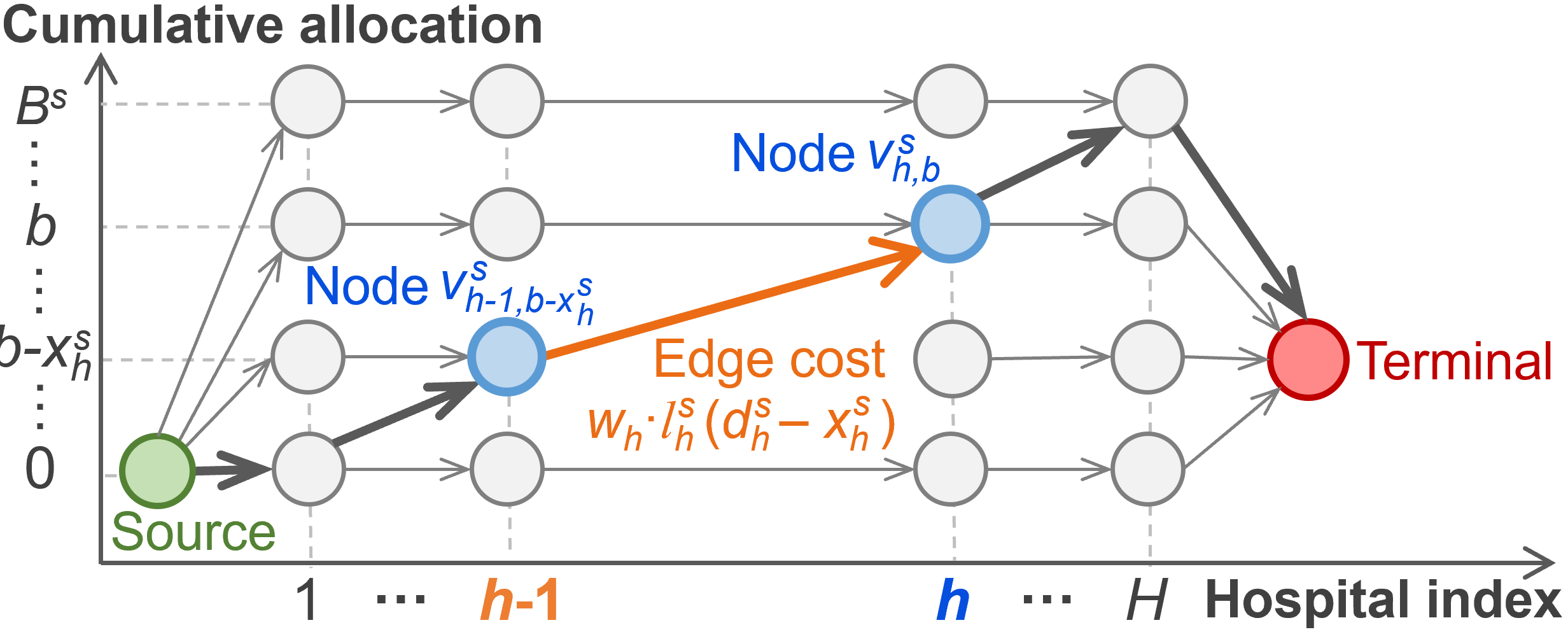}
\vspace{-18pt}
\caption{Layered-DAG reformulation.}
\vspace{-15pt}
\label{fig:DAG}
\end{wrapfigure}
As illustrated in Fig.~\ref{fig:DAG}, each supply-specific subproblem is represented by a layered DAG.
For supply type $s$, layer $h$ corresponds to hospital $h$, while node $v_{h,b}^s$ represents a cumulative allocation of $b$ units after the first $h$ hospitals.
An edge from $v_{h-1,b-x_h^s}^s$ to $v_{h,b}^s$ represents assigning $x_h^s$ units to hospital $h$ and incurs cost $w_h \cdot \ell_h^s(d_h^s-x_h^s)$.
Thus, each source-to-terminal path corresponds to a feasible allocation, and the minimum-cost path yields the optimal solution to formulation~(\ref{eq:allocation}).
Repeating this construction across supply types yields the complete allocation $\{x_h^s\}$.
The transformation changes the apparent operation from resource allocation to layered-DAG shortest-path search while preserving a reliable mapping back to the original solution.
Compared with the initial mathematical reformulation, this alternative formulation further obscures task intent, but its mathematical structure may still leave some clues, as reflected by the \emph{Math-only} results in Table~\ref{tab:main-results}.

\textbf{Semantic rewriting.}
\texttt{PriCon} therefore recasts the path-search structure as a network-routing story before cloud consultation:

\begin{tcolorbox}[
    breakable,
    colback=myorange!5,
    colframe=myorange,
    boxrule=0.7pt,
    arc=3mm,
    left=4pt,
    right=4pt,
    top=3pt,
    bottom=3pt
]

\setlength{\leftskip}{0.3em}
\setlength{\rightskip}{0.3em}
\small\ttfamily
\linespread{1.1}\selectfont
``%
Given multiple \context{layered communication networks} [N], each with \context{network states} [V], \context{available links} [E], and associated \context{transmission costs} [C],
find a \operation{minimum-cost route through each network} from its source gateway to a destination gateway.%
''
\end{tcolorbox}

Under this interpretation, graph states, transitions, and costs become network states, communication links, and transmission costs.
The cloud therefore reasons over a network-routing task rather than the original medical-supply allocation task, further obscuring its context and operation.

Appendix~\ref{app:rewrite-details} provides detailed local correspondences $\mathcal{M}$ and representative cloud outputs for the medical-supply example.
Appendix~\ref{app:logic-recipe} further presents an interesting \texttt{PriCon} case transforming logic-grid reasoning into cooking-procedure planning.

\subsection{Local Review and Closed-Loop Refinement}
\label{sec:privacy-review}

\begin{figure}[t]
\centering\includegraphics[width=0.9\linewidth]{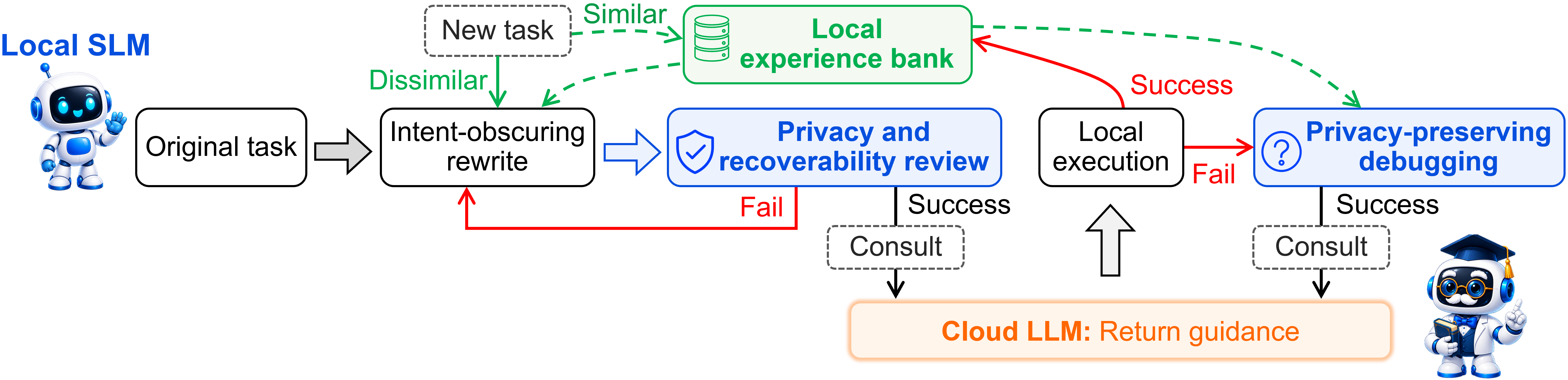}
\vspace{-8pt}
\caption{Closed-loop consultation and refinement in \texttt{PriCon}.}
\vspace{-12pt}
\label{fig:PriCon-system}
\end{figure}

\vspace{-2pt}
Stronger transformations can better obscure task intent but make cloud guidance harder to map back for local execution.
Preserving more of the original structure improves recoverability but may retain intent-revealing cues.
As shown in Fig.~\ref{fig:PriCon-system}, \texttt{PriCon} therefore forms a closed refinement loop: failed reviews trigger further rewriting, execution failures trigger privacy-preserving debugging, and successful executions are stored in a local experience bank.

\vspace{-3pt}
\begin{enumerate}[label=\arabic*), leftmargin=1.5em, itemsep=0.5pt, topsep=2pt]
\item
\textbf{Privacy and recoverability review.}
Before transmission, a local reviewer checks whether sensitive values $\mathcal{V}$ are exposed, whether the protected intent is inferable from semantic or structural cues, and whether $\mathcal{M}$ can map cloud guidance back to the original task.
If any check fails, the SLM revises the surrogate and repeats the review within a refinement budget.

\item
\textbf{Privacy-preserving debugging.}
Cloud guidance may still be too abstract or complex for the local SLM to execute reliably.
Instead of returning task-specific failures to the cloud, the SLM abstracts them into surrogate-level diagnostic requests and subjects them to the same privacy and recoverability review.
For example, it may ask the cloud to decompose a difficult surrogate reasoning step without revealing the original task or local execution trace.

\item
\textbf{Local experience bank.}
Successful consultations are stored locally as reusable experiences, including the mathematical representation, optional alternative formulation, semantic rewrite, correspondence $\mathcal{M}$, and validated refinement or debugging trajectories.
When a new task arrives, \texttt{PriCon} retrieves prior experiences by comparing their mathematical structures rather than rewritten stories, using four fields: problem type, decision variables, objective, and constraints.
Prior experiences are ranked by embedding similarity across these fields; see Appendix~\ref{app:experience-retrieval} for details.
If a sufficiently similar experience is found, it is used to initialize surrogate construction, reducing LLM consultation rounds and token usage; otherwise, \texttt{PriCon} constructs a new surrogate and stores its validated trajectory as a new experience.
\end{enumerate}

\vspace{-2pt}
\section{Experimental Results}
\label{sec:results}

\vspace{-2pt}
\subsection{Experimental Setup}

\vspace{-2pt}
\textbf{Models and tasks.}
We evaluate \texttt{PriCon} with three models serving as local-side SLMs: GPT-4o-mini, DeepSeek-V4-Flash, and Qwen3.5-Flash, each paired with DeepSeek-V4 Pro as the cloud LLM~\citep{openai2024gpt4omini,deepseek2026v4,qwen2026qwen35}.
Our benchmark contains 10 task families spanning structured and long-context reasoning, decision and optimization, coding, and cross-domain composite tasks, as summarized in Table~\ref{tab:task-families}.
We retain only tasks that local SLMs cannot reliably solve end-to-end, thus requiring cloud LLM guidance; see Appendix~\ref{app:tasks} for the screening procedure.
After screening, we retain 10 diverse instances per family, yielding \emph{100 tasks in total}.
All retained task instances are available in our GitHub repository.%
\footnote{ \url{https://github.com/guantian111/Pricon_tasks}}

\textbf{Comparison settings.}
We compare direct cloud consultation, sensitive-value removal, a decoy-based obfuscation baseline~\citep{huang2026homellama}, and \texttt{PriCon}.
For \texttt{PriCon}, we evaluate the complete semantic rewriting process under context and joint protection.
We additionally report a \emph{Math-only} ablation that stops before semantic rewriting, isolating its benefit.
Operation-only protection is evaluated separately to study context-operation coupling.
For the decoy baseline, we use 5 and 20 decoys to examine its privacy-overhead trade-off.

\textbf{Metrics.}
As each benchmark task may contain multiple subtasks, we evaluate privacy and utility at the subtask level.
For privacy, we ask the cloud LLM to infer three targets from the complete cloud-visible transcript: task context, task operation, and their pair (joint inference).
For each task, \emph{Hit@1} and \emph{Hit@5} measure the proportions of its subtasks whose ground-truth targets appear among the cloud LLM's top-1 and top-5 guesses, respectively;
joint inference requires the correct context-operation pair.
For utility, we compute each task's subtask success rate as the proportion of its subtasks yielding functionally correct final solutions, with direct cloud consultation as the reference.
In the main results, we average these per-task privacy and utility scores across all 100 benchmark tasks.
We additionally report LLM token usage and consultation rounds as system overhead.

\vspace{-3pt}
\subsection{Main Results}
\label{sec:main-results}
\begin{table}[!t]
\centering
\captionsetup{font=scriptsize,skip=2pt}
\caption{Task families in our evaluation benchmark.}
\label{tab:task-families}
\scriptsize
\setlength{\tabcolsep}{3pt}
\begin{tabular}{p{0.3\linewidth}p{0.65\linewidth}}
\toprule
\textbf{Category} & \textbf{Task families} \\
\midrule
\textbf{Structured and long-context reasoning}
& Alibi verification, logic grid, counterfactual reasoning, recursive expansion, relation graph \\
\textbf{Decision and optimization}
& Financial modeling, budget allocation, office setup \\
\textbf{Coding}
& Python coding \\
\textbf{Cross-domain composite}
& Composite cross-domain \\
\bottomrule
\end{tabular}
\vspace{-5pt}
\end{table}

\noindent\textbf{Task-intent protection.}
We first evaluate task-intent protection across the three local SLMs, with the local experience bank disabled to isolate the effect of rewriting.
Table~\ref{tab:main-results} compares the cloud LLM's task-intent inference under different consultation settings, with detailed category-wise results in Appendix~\ref{app:category-results}.
Direct cloud consultation yields Hit@1 scores of 94--99\%, while sensitive-value removal yields similarly high scores of 93--99\%, showing that hiding sensitive values alone provides little protection against task-intent inference.
The decoy baseline reduces inference, but operation Hit@1 remains 18--30\% even with 20 decoys.
In contrast, \texttt{PriCon-Context} with semantic rewriting reduces context Hit@1 to 0\% across all three local SLMs.
Under joint protection, \texttt{PriCon-Joint} with semantic rewriting achieves 0\% Hit@1 for all three inference targets, while Hit@5 remains at 0\% in nearly all cases.

\noindent\textbf{Effect of semantic rewriting.}
The \emph{Math-only} ablation already substantially reduces inference of the protected intent, showing that the mathematical stages provide strong protection.
Under context protection, its context Hit@1 is 1--3\%, whereas semantic rewriting reduces it to 0\% across all three local SLMs.
Under joint protection, some inference remains with \emph{Math-only}, whereas semantic rewriting reduces Hit@1 to 0\% across all inference targets.
These results support using mathematical reformulation as an intermediate stage and semantic rewriting as the final cloud-facing form. 

\begin{table*}[!t]
\centering
\captionsetup{justification=centering,font=scriptsize,skip=2pt}
\caption{Task-intent inference by the cloud LLM under different consultation settings and local SLMs. \\
Entries report task-averaged Hit@1 / Hit@5 (\%) over 100 benchmark tasks.}
\label{tab:main-results}
\tiny
\setlength{\aboverulesep}{0.5pt}
\setlength{\belowrulesep}{0.5pt}
\begin{tabular}{cc|cccccccc}
\toprule
\multirow{2}{*}{\makecell{\textbf{Local SLM}}} &
\multirow{2}{*}{\makecell{\textbf{Inference}\\\textbf{Target}}} &
\multirow{2}{*}{\makecell{Direct cloud\\consultation}} &
\multirow{2}{*}{\makecell{Sensitive-value\\removal}} &
\multicolumn{2}{c}{Decoy} &
\multicolumn{2}{c}{\texttt{PriCon-Context}} &
\multicolumn{2}{c}{\texttt{PriCon-Joint}} \\
\cmidrule(lr){5-6}
\cmidrule(lr){7-8}
\cmidrule(lr){9-10}
& & & & 5 & 20 &
Math-only & \textbf{Semantic} &
Math-only & \textbf{Semantic} \\
\midrule
\multirow{3}{*}{\makecell{GPT-4o-mini}}
& \bluecell{Context} & \bluecell{94 / 98} & \bluecell{94 / 98} & \bluecell{10 / 32} & \bluecell{6 / 21} & \bluecell{1 / 3} & \bluecell{\textbf{0 / 0}} & \bluecell{3 / 10} & \bluecell{\textbf{0 / 0}} \\
& Operation & 98 / 100 & 99 / 100 & 30 / 65 & 18 / 50 & 49 / 63 & 31 / 42 & 1 / 5 & \textbf{0 / 0} \\
& \cellcolor{orange!10}Joint & \cellcolor{orange!10}95 / 98 & \cellcolor{orange!10}94 / 98 & \cellcolor{orange!10}7 / 31 & \cellcolor{orange!10}3 / 13 & \cellcolor{orange!10}0 / 0 & \cellcolor{orange!10}\textbf{0 / 0} & \cellcolor{orange!10}0 / 1 & \cellcolor{orange!10}\textbf{0 / 0} \\
\midrule
\multirow{3}{*}{\makecell{DeepSeek-V4-Flash}}
& \bluecell{Context} & \bluecell{98 / 100} & \bluecell{96 / 99} & \bluecell{9 / 34} & \bluecell{8 / 20} & \bluecell{3 / 3} & \bluecell{\textbf{0 / 0}} & \bluecell{1 / 4} & \bluecell{\textbf{0 / 0}} \\
& Operation & 95 / 99 & 95 / 99 & 26 / 63 & 30 / 68 & 48 / 62 & 59 / 75 & 1 / 4 & \textbf{0 / 0} \\
& \cellcolor{orange!10}Joint & \cellcolor{orange!10}98 / 100 & \cellcolor{orange!10}96 / 99 & \cellcolor{orange!10}6 / 30 & \cellcolor{orange!10}4 / 15 & \cellcolor{orange!10}0 / 1 & \cellcolor{orange!10}\textbf{0 / 0} & \cellcolor{orange!10}0 / 2 & \cellcolor{orange!10}\textbf{0 / 0} \\
\midrule
\multirow{3}{*}{\makecell{Qwen3.5-Flash}}
& \bluecell{Context} & \bluecell{95 / 97} & \bluecell{93 / 99} & \bluecell{10 / 38} & \bluecell{3 / 18} & \bluecell{1 / 3} & \bluecell{\textbf{0 / 0}} & \bluecell{2 / 6} & \bluecell{\textbf{0 / 0}} \\
& Operation & 99 / 100 & 96 / 98 & 29 / 66 & 18 / 46 & 48 / 64 & 56 / 69 & 2 / 7 & \textbf{0 / 2} \\
& \cellcolor{orange!10}Joint & \cellcolor{orange!10}96 / 97 & \cellcolor{orange!10}93 / 99 & \cellcolor{orange!10}7 / 30 & \cellcolor{orange!10}4 / 14 & \cellcolor{orange!10}0 / 0 & \cellcolor{orange!10}\textbf{0 / 0} & \cellcolor{orange!10}0 / 2 & \cellcolor{orange!10}\textbf{0 / 0} \\
\bottomrule
\end{tabular}
\vspace{-5pt}
\end{table*}

\noindent\textbf{Utility.}
Table~\ref{tab:utility-results} shows that the complete \texttt{PriCon} with semantic rewriting achieves 100\% subtask success across all three local SLMs, matching direct cloud consultation.
Thus, task-intent protection does not reduce cloud-assisted utility on our benchmark.

\begin{table*}[!t]
\centering
\captionsetup{justification=centering,font=scriptsize,skip=2pt}
\caption{Cloud-assisted utility under different consultation settings.\\
Entries report task-averaged subtask success rate (\%) over 100 benchmark tasks.}
\label{tab:utility-results}
\tiny
\setlength{\aboverulesep}{0.5pt}
\setlength{\belowrulesep}{0.5pt}
\begin{tabular}{c|cccccccc}
\toprule
\multirow{2}{*}{\makecell{\textbf{Local SLM}}} &
\multirow{2}{*}{\makecell{Direct cloud\\consultation}} &
\multirow{2}{*}{\makecell{Sensitive-value\\removal}} &
\multicolumn{2}{c}{Decoy} &
\multicolumn{2}{c}{\texttt{PriCon-Context}} &
\multicolumn{2}{c}{\texttt{PriCon-Joint}} \\
\cmidrule(lr){4-5}
\cmidrule(lr){6-7}
\cmidrule(lr){8-9}
& & & 5 & 20 &
Math-only & \textbf{Semantic} &
Math-only & \textbf{Semantic} \\
\midrule
GPT-4o-mini
& 100 & 100 & 100 & 100 & 100 & 100 & 100 & 100 \\
\midrule
DeepSeek-V4-Flash
& 100 & 100 & 100 & 100 & 100 & 100 & 100 & 100 \\
\midrule
Qwen3.5-Flash
& 100 & 100 & 100 & 100 & 100 & 100 & 100 & 100 \\
\bottomrule
\end{tabular}
\vspace{-10pt}
\end{table*}

\noindent\textbf{Consultation overhead.}
Direct cloud consultation and sensitive-value removal each require only one round: the former sends the task directly, while the latter simply replaces sensitive values with placeholders.
However, direct consultation sends all task-specific inputs to the cloud LLM, so its token cost can grow substantially with input size.
\texttt{PriCon} introduces additional rewriting and refinement, requiring 5.6--7.2 consultation rounds and 1.96--2.52M LLM tokens with semantic rewriting.
Nevertheless, this overhead remains far below that of the 20-decoy setting, which requires 35.9--39.8 rounds and 11.4--12.7M tokens.
Together with Table~\ref{tab:main-results}, these results show that \texttt{PriCon} achieves stronger task-intent protection at much lower overhead than heavy decoy-based obfuscation. 

\begin{table*}[!t]
\centering
\captionsetup{font=scriptsize,skip=2pt}
\vspace{-6pt}
\caption{Average consultation overhead across 100 benchmark tasks.
LLM token usage is reported in millions (M = $10^6$).}
\label{tab:consultation-overhead}
\tiny
\setlength{\aboverulesep}{0.5pt}
\setlength{\belowrulesep}{0.5pt}
\begin{tabular}{cc|cccccccc}
\toprule
\multirow{2}{*}{\makecell{\textbf{Local SLM}}} &
\multirow{2}{*}{\textbf{Metric}} &
\multirow{2}{*}{\makecell{Direct cloud\\consultation}} &
\multirow{2}{*}{\makecell{Sensitive-value\\removal}} &
\multicolumn{2}{c}{Decoy} &
\multicolumn{2}{c}{\texttt{PriCon-Context}} &
\multicolumn{2}{c}{\texttt{PriCon-Joint}} \\
\cmidrule(lr){5-6}
\cmidrule(lr){7-8}
\cmidrule(lr){9-10}
& & & & 5 & 20 &
Math-only & \textbf{Semantic} &
Math-only & \textbf{Semantic} \\
\midrule
\multirow{2}{*}{\makecell{GPT-4o-mini}}
& LLM token usage     & {0.49 M} & {0.34 M} & 2.48 M & 11.4 M & 2.29 M & 1.96 M & 2.39 M & 2.52 M \\
& Consultation rounds & 1 & 1 & 7.8 & 35.9 & 7.2 & 5.6 & 7.5 & 7.2  \\
\midrule
\multirow{2}{*}{\makecell{DeepSeek-V4-Flash}}
& LLM token usage     & {0.65 M} & {0.35 M} & 2.86 M & 12.1 M & 2.29 M & 2.03 M & 2.39 M & 2.31 M \\
& Consultation rounds & 1 & 1 & 9.0 & 38.1 & 7.2 & 7.1 & 5.8 & 6.6 \\
\midrule
\multirow{2}{*}{\makecell{Qwen3.5-Flash}}
& LLM token usage     & {0.54 M} & {0.39 M} & 2.86 M & 12.7 M & 2.51 M & 2.0 M & 2.60 M & 2.52 M \\
& Consultation rounds & 1 & 1 & 9.0 & 39.8 & 7.9 & 5.7 & 7.7 & 7.2 \\
\bottomrule
\end{tabular}
\vspace{-5pt}
\end{table*}

\noindent\textbf{Context-operation coupling.}
Although task context and operation are distinct privacy targets, contextual cues may still reveal the protected operation.
With operation-only protection, operation Hit@1 remains at 2--3\% when the original context is retained, whereas joint protection reduces it to 0\% across all local SLMs.
These results show that task context can act as a side channel for operation inference, supporting \texttt{PriCon}'s strategy of transforming context whenever the operation is protected.

\begin{table*}[!t]
\centering
\captionsetup{font=scriptsize,skip=2pt}
\caption{Context-operation coupling under operation-only and joint protection.
Entries report Hit@1 / Hit@5 (\%).}
\label{tab:context-operation-coupling}
\tiny
\setlength{\aboverulesep}{0.5pt}
\setlength{\belowrulesep}{0.5pt}

\begin{tabular}{cc|ccccc}
\toprule
\multirow{2}{*}{\textbf{Local SLM}} &
\multirow{2}{*}{\makecell{\textbf{Inference Target}}} &
\multirow{2}{*}{\makecell{Sensitive-value\\removal}} &
\multicolumn{2}{c}{\texttt{Operation-only}} &
\multicolumn{2}{c}{\texttt{PriCon-Joint}} \\
\cmidrule(lr){4-5}
\cmidrule(lr){6-7}
& & &
Math-only & Semantic &
Math-only & \textbf{Semantic} \\
\midrule
\multirow{2}{*}{\makecell{GPT-4o-mini}}
& \greencell{Operation} & \greencell{99 / 100} & \greencell{3 / 10} & \greencell{3 / 6} & \greencell{1 / 5} & \greencell{\textbf{0 / 0}} \\
& Joint & 94 / 98  & 2 / 5 & { 1 / 5}   & 0 / 1   &  \textbf{0 / 0} \\
\midrule
\multirow{2}{*}{\makecell{DeepSeek-V4-Flash}}
& \greencell{Operation} & \greencell{95 / 99} & \greencell{1 / 7} & \greencell{2 / 4} & \greencell{1 / 4} & \greencell{\textbf{0 / 0}} \\
& Joint & 96 / 99 & 1 / 4 & {1 / 4}   & 0 / 2   & \textbf{0 / 0} \\
\midrule
\multirow{2}{*}{\makecell{Qwen3.5-Flash}}
& \greencell{Operation} & \greencell{96 / 98} & \greencell{3 / 9} & \greencell{2 / 7} & \greencell{2 / 7} & \greencell{\textbf{0 / 2}} \\
& Joint & 93 / 99 & 0 / 4 & {0 / 2}   & 0 / 2  & \textbf{0 / 0} \\
\bottomrule
\end{tabular}
\vspace{-5pt}
\end{table*}

\begin{figure*}[!t]
\centering
\subfigure[\scriptsize LLM token usage ratio.]{
\includegraphics[width=0.48\linewidth]{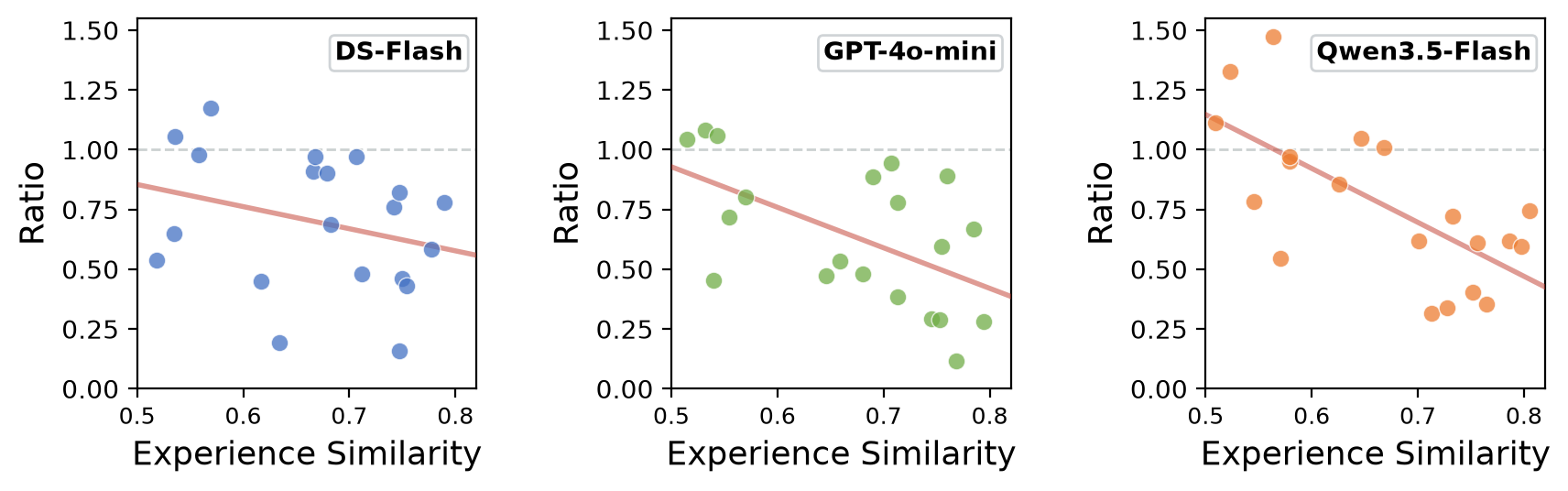}}
\hspace{0.1cm}
\subfigure[\scriptsize Consultation round ratio.]{
\includegraphics[width=0.48\linewidth]{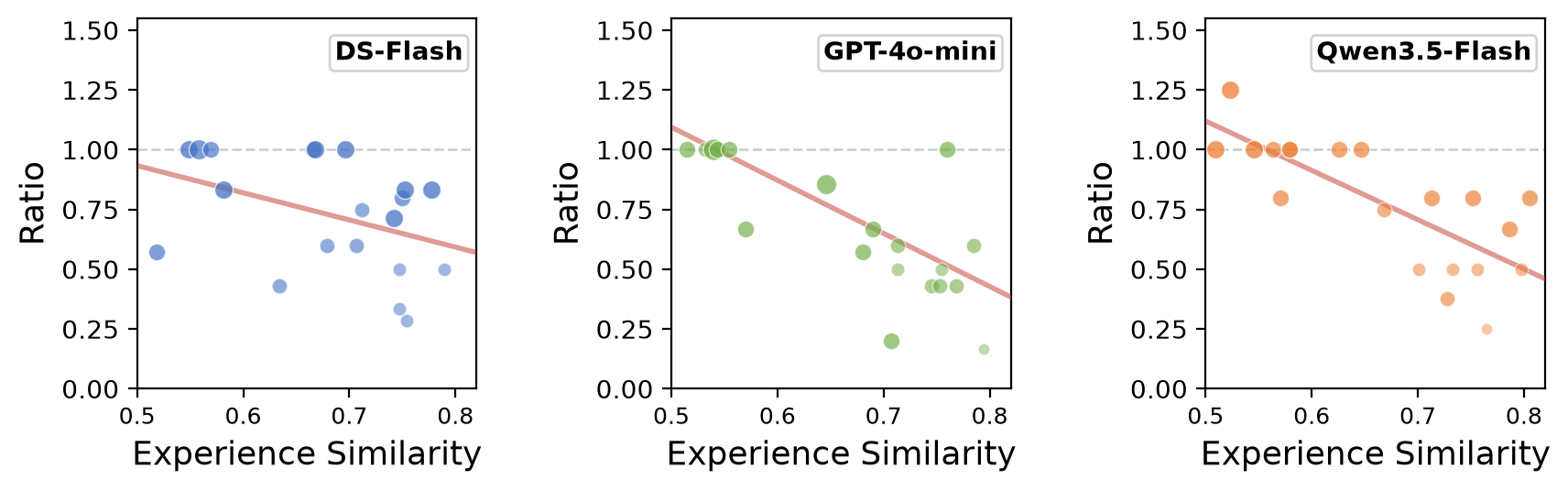}}
\vspace{-12pt}
\captionsetup{justification=centering,font=scriptsize}
\caption{Effect of experience similarity on consultation overhead under PriCon-Joint with semantic rewriting. \\
Each ratio is computed as the overhead with experience reuse divided by that without experience reuse.}
\vspace{-13pt}
\label{fig:experience-bank}
\end{figure*}

\noindent\textbf{Experience reuse.}
We finally evaluate experience reuse on held-out tasks.
Figure~\ref{fig:experience-bank} shows that both LLM token usage and consultation rounds generally decrease as experience similarity increases.
At low similarity (below approximately 0.55), retrieved experiences may introduce irrelevant information and increase token usage.
For consultation rounds, some tasks also show little reduction because their original overhead is already low.
For more demanding cases, reuse can reduce consultation rounds to approximately 20\% of the baseline.
These observations suggest using a similarity threshold around 0.6 to avoid ineffective reuse; for less similar tasks, \texttt{PriCon} constructs surrogates independently and stores them as new experiences for future reuse.

\vspace{-4pt}
\section{Discussion}
\vspace{-4pt}

\textbf{Applicability boundary.}
\texttt{PriCon} is most suitable when the task context can be separated from the underlying task operation needed for cloud reasoning.
It is less applicable to knowledge queries whose answers are inherently tied to the task context, such as the effects or use of a particular medication.
Such queries may instead be handled by a local knowledge base without cloud consultation.
Operation protection is also challenging for complex tasks, where finding an alternative formulation with a reliable solution mapping can be difficult.
A richer experience bank may expand this boundary.

\textbf{A moving capability frontier.}
As local SLMs become more capable, more tasks will be completed entirely on-device.
Task-private consultation will therefore shift toward harder reasoning, planning, and decision tasks that still benefit from cloud guidance.
Advances in local SLMs and cloud LLMs may also enable more complex transformations, further extending this frontier.

\textbf{Beyond language models.}
This work focuses on text-based SLM-LLM consultation, while task intent may also be revealed through images, audio, and other modalities.
Extending \texttt{PriCon} to multimodal models may require modality-specific transformations of objects, scenes, or spatial relations while preserving information needed for reasoning.
Extending task-private consultation beyond text is an important direction for multimodal local-cloud collaboration.

\newpage

\subsubsection*{Acknowledgments}
We thank Mingcai Chen, Qiao Zhang, and Shuai Wang for their valuable feedback on this work.

\bibliography{iclr2027_conference}
\bibliographystyle{iclr2027_conference}

\newpage
\appendix

\section*{\textbf{Appendices}}

\section{Local Correspondences for the Medical-Supply Example}
\label{app:rewrite-details}
We provide detailed examples of how \texttt{PriCon} transforms the medical-supply task into a semantic surrogate and maps cloud guidance back to the original task.
We show the necessary intermediate mathematical representations, followed by semantic rewriting, representative cloud guidance, and local execution.

\subsection{Context Protection}
\begin{center}
\small
\begin{tabular}{ll}
\toprule
Original task & Intermediate mathematical representation \\
\midrule
Hospital $h$ & Demand group $g_h$ \\
Supply type $s$ & Resource type $r_s$ \\
Priority $w_h$ & Priority weight $w_{g_h}$ \\
Available stock $B^s$ & Resource capacity $B^{r_s}$ \\
Demand $d_h^s$ & Resource requirement $d_{g_h}^{r_s}$ \\
Allocation $x_h^s$ & Resource allocation $x_{g_h}^{r_s}$ \\
Shortage impact $\ell_h^s(\cdot)$ & Unmet-requirement impact $\ell_{g_h}^{r_s}(\cdot)$ \\
\bottomrule
\end{tabular}
\end{center}
This intermediate representation preserves the allocation structure needed for subsequent rewriting and local recovery, while the concrete values remain local.

\textbf{Semantic rewriting and local correspondence $\mathcal{M}$.}
\texttt{PriCon} then rewrites the intermediate mathematical representation as a GPU-allocation task.
The correspondence below allows cloud guidance expressed in GPU terminology to be mapped back through the mathematical representation to the original healthcare task.
\begin{center}
\small
\begin{tabular}{ll}
\toprule
Intermediate mathematical representation & Semantic surrogate \\
\midrule
Demand group $g_h$ & Model-training job $j_h$ \\
Resource type $r_s$ & GPU resource type $u_s$ \\
Priority weight $w_{g_h}$ & Job priority $w_{j_h}$ \\
Resource capacity $B^{r_s}$ & GPU capacity $B^{u_s}$ \\
Resource requirement $d_{g_h}^{r_s}$ & Compute demand $d_{j_h}^{u_s}$ \\
Resource allocation $x_{g_h}^{r_s}$ & GPU allocation $x_{j_h}^{u_s}$ \\
Unmet-requirement impact $\ell_{g_h}^{r_s}(\cdot)$ & Unmet-compute impact $\ell_{j_h}^{u_s}(\cdot)$ \\
\bottomrule
\end{tabular}
\end{center}

\textbf{LLM guidance.}
The cloud reasons entirely over the rewritten GPU-allocation task.
It processes each GPU type independently, considers feasible GPU allocations for each training job, and identifies the allocation minimizing cumulative unmet-compute impact. 
A representative \emph{reasoning response} is:
\begin{itemize}[leftmargin=1.5em, itemsep=0pt, topsep=2pt]
    \item Process each GPU resource type independently because their capacities are separate.
    \item For one GPU type, consider the training jobs while tracking the capacity already allocated.
    \item For each job, enumerate feasible GPU allocations up to its compute demand and the remaining capacity.
    \item Update the cumulative unmet-compute impact and retain the lowest-cost feasible allocation.
    \item Repeat the procedure for all GPU resource types.
\end{itemize}

Alternatively, the cloud may return code implementing the same GPU-allocation procedure:
\begin{verbatim}
def allocate_gpu(job_demand, job_priority, gpu_capacity, loss):
    ...
    limit = min(job_demand[i], gpu_capacity - used)
    ...
    cost = dp[i][used] + job_priority[i] * loss(i, job_demand[i] - x)
    ...
    return gpu_allocation
\end{verbatim}
Here, \texttt{job\_demand}, \texttt{job\_priority}, and \texttt{gpu\_capacity} correspond to the demand, weight, and capacity arguments in the mathematical surrogate.

\textbf{Local execution.}
Consider one supply type with local demands $d^s=[8,5,7]$, priority weights $w=[0.5,0.3,0.2]$, and available stock $B^s=10$.
The SLM instantiates the cloud guidance with these local values and obtains the surrogate allocation $[6,2,2]$.
Through $\mathcal{M}$, model-training jobs are mapped back to demand groups and hospitals, while GPU resources are mapped back to medical-supply types.
The allocation is therefore recovered locally as $[6,2,2]$ for the corresponding hospitals.

\subsection{Operation and Joint Protection}

\textbf{Intermediate alternative formulation.}
When the task operation is protected, \texttt{PriCon} further transforms each supply-specific allocation problem into a layered-DAG shortest-path problem.
\begin{center}
\small
\begin{tabular}{ll}
\toprule
Allocation representation & Intermediate alternative mathematical representation \\
\midrule
Resource type $r_s$ & Layered graph $G^s$ \\
Demand group $g_h$ & Layer $h$ \\
Cumulative allocation $b$ & State $v_{h,b}^s$ \\
Resource allocation $x_{g_h}^{r_s}$ & Transition increment $x_{g_h}^{r_s}$ \\
Feasible allocation choice & Directed edge \\
Unmet-requirement impact & Edge cost \\
Initial allocation state & Source node $v_{0,0}^s$ \\
Complete allocation state & Terminal node $v_{H,b}^s$ \\
Allocation sequence & Source-to-terminal path \\
Optimal allocation & Minimum-cost path \\
\bottomrule
\end{tabular}
\end{center}

\textbf{Semantic rewriting and local correspondence $\mathcal{M}$.}
\texttt{PriCon} then rewrites the layered-DAG representation as a communication-network routing task before cloud consultation.
The correspondence below maps a returned network route first to the intermediate DAG path and then to the original allocation.
\begin{center}
\small
\begin{tabular}{ll}
\toprule
Intermediate alternative mathematical representation & Semantic surrogate \\
\midrule
Layered graph $G^s$ & Layered communication network $N^s$ \\
Layer $h$ & Network layer $h$ \\
State $v_{h,b}^s$ & Network state $q_{h,b}^s$ \\
Transition increment $x_{g_h}^{r_s}$ & Routed resource increment \\
Directed edge & Available communication link \\
Edge cost & Transmission cost \\
Source node $v_{0,0}^s$ & Source gateway \\
Terminal node $v_{H,b}^s$ & Destination gateway \\
Source-to-terminal path & End-to-end route \\
Minimum-cost path & Minimum-cost route \\
\bottomrule
\end{tabular}
\end{center}

Each source-to-terminal path therefore encodes a feasible allocation, allowing the local SLM to recover the original allocation from the increments between consecutive states.

\textbf{LLM guidance.}
The cloud now describes the same shortest-path reasoning in communication-network terminology: it propagates minimum transmission costs across network layers, updates reachable states through available links, records predecessors, and backtracks from the minimum-cost destination.
A representative \emph{reasoning response} is:
\begin{itemize}[leftmargin=1.5em, itemsep=0pt, topsep=2pt]
    \item Traverse the layered communication network from the source gateway toward the destination layer.
    \item Maintain the minimum cumulative transmission cost of reaching each network state.
    \item Update reachable states through available links and record the predecessor whenever a lower-cost route is found.
    \item At the destination layer, select the gateway with the minimum cumulative cost and backtrack to recover the route.
    \item Repeat the procedure independently for each network.
\end{itemize}

The cloud may also return code implementing the same routing procedure:

\begin{verbatim}
def min_cost_route(layers, links, source_gateway,
                   destination_states):
    ...
    for v, transmission_cost in links.get(u, []):
        cand = cost[u] + transmission_cost
        ...
    route.reverse()
    return route, cost[destination]
\end{verbatim}

Here, \texttt{links}, \texttt{transmission\_cost}, and \texttt{route} replace the corresponding graph edges, edge costs, and path in the alternative mathematical formulation.

\textbf{Local execution.}
The local SLM first maps the returned network route through $\mathcal{M}$ to the layered-DAG path
$v_{0,0}^s \rightarrow v_{1,6}^s \rightarrow v_{2,8}^s \rightarrow v_{3,10}^s$.
It then recovers the allocation from the state increments as $x^s=[6,2,2]$.
Thus, the cloud interacts entirely with the network-routing task, while the local correspondence maps its guidance through the intermediate DAG representation back to the original allocation.

\section{Additional Case Study: From Logic-Grid Reasoning to Recipe Planning}
\label{app:logic-recipe}

This section provides an additional example illustrating the rewriting process in \texttt{PriCon}.
Unlike the medical-supply example, the original task here requires multi-stage logic-grid reasoning.
\texttt{PriCon} first captures its reasoning structure in an intermediate mathematical representation and, when the operation is protected, further transforms it into layered state-transition planning.
Semantic rewriting then recasts this structure as a different story for cloud consultation.

\subsection{Original Task}

We begin with the following logic-grid reasoning task.

\begin{tcolorbox}[
    breakable,
    colback=black!2,
    colframe=black!45,
    boxrule=0.6pt,
    arc=3mm,
    left=4pt,
    right=4pt,
    top=3pt,
    bottom=3pt
]
\small\ttfamily
\linespread{1.03}\selectfont

Please complete the following four-stage logical grid reasoning task in order.

\medskip
\textbf{=== Background ===}

There are 5 people (Zhang, Wang, Li, Zhao, Chen), each living on a different floor (floors 1--5), each with a different pet (dog/cat/bird/fish/rabbit) and a different occupation (doctor/teacher/lawyer/engineer/accountant).

The following 6 clues are known:

1. Zhang lives on a floor above Li (not necessarily adjacent).

2. The doctor lives on floor 3.

3. Wang's pet is a dog.

4. Zhao's occupation is teacher.

5. The person who keeps a cat lives on a floor above the person who keeps a bird.

6. Chen's occupation is neither engineer nor accountant.

\medskip
\textbf{=== Stage 1: Initial Extraction ===}

Extract all directly knowable facts from the clues (no inference required).

For example: ``Doctor = floor 3'', ``Wang's pet = dog'' are direct facts; ``Zhang is above Li'' is a relationship.

\medskip
\textbf{=== Stage 2: First Round of Inference (using Stage 1 results) ===}

Use Clue 1 (Zhang above Li) and Clue 5 (cat above bird):

- Given the floor relationship between Zhang and Li, combined with the 5-floor constraint, which floors can they each be on?

- What does the floor relationship between cat and bird imply? Combine with Clue 3 (Wang = dog).

List the possible floor ranges for the 5 people.

\medskip
\textbf{=== Stage 3: Second Round of Inference (using Stage 1--2 results) ===}

Use Clue 2 (doctor = floor 3), Clue 4 (Zhao = teacher), Clue 6 (Chen is not engineer/accountant):

- After the doctor's floor is determined, what restrictions does this place on the others?

- Which possibilities are narrowed down by the occupational constraints of Zhao and Chen?

Combined with Stage 2, what assignments can currently be determined?

\medskip
\textbf{=== Stage 4: Final Inference ===}

a) Output the final answer: a list of 5-tuples
[(Name, Floor, Pet, Occupation), \ldots]

b) Verification: Is each clue satisfied? List the verification result for each clue.

c) What pet does the engineer keep?

\end{tcolorbox}

\subsection{Context Protection}

\textbf{Intermediate mathematical representation.}
\texttt{PriCon} first abstracts the logic-grid task into a constraint-reasoning problem in which people become entities, floors become ordered positions, pets become attributes, and occupations become categories.
The original ordering, assignment, and exclusion relations are preserved in this representation.

\begin{tcolorbox}[
    breakable,
    colback=myblue!5,
    colframe=myblue,
    boxrule=0.7pt,
    arc=3mm,
    left=4pt,
    right=4pt,
    top=3pt,
    bottom=3pt
]
\small\ttfamily
\linespread{1.03}\selectfont

Please complete the following four-stage constraint-reasoning task.

\medskip
\textbf{=== Background ===}

There are [N] entities
$\{e_1,e_2,\ldots,e_N\}$.
Each entity is assigned a distinct ordered position from
$\{p_1,p_2,\ldots,p_N\}$,
one distinct value from attribute set
$\mathcal{A}=\{a_1,\ldots,a_N\}$,
and one distinct value from category set
$\mathcal{B}=\{b_1,\ldots,b_N\}$.

The following constraints are known:

1. $e_1$ is assigned a higher position than $e_3$.

2. The entity with category $b_1$ occupies position $p_{[K]}$.

3. Entity $e_2$ has attribute $a_1$.

4. Entity $e_4$ has category $b_2$.

5. The entity with attribute $a_2$ occupies a higher position than the entity with attribute $a_3$.

6. Entity $e_5$ has neither category $b_3$ nor $b_4$.

\medskip
\textbf{=== Stage 1: Initial Extraction ===}

Extract all constraints that can be read directly from the specification without additional inference.

For example, ``$b_1=p_{[K]}$'' and ``$e_2=a_1$'' are direct assignments, while ``$e_1$ is above $e_3$'' is an ordering relation.

\medskip
\textbf{=== Stage 2: First Round of Inference ===}

Use the ordering constraints involving $(e_1,e_3)$ and $(a_2,a_3)$:

- Determine the feasible position ranges of $e_1$ and $e_3$.

- Determine the feasible relative positions of $a_2$ and $a_3$.

- Combine these relations with $e_2=a_1$.

List the remaining feasible position ranges for all entities.

\medskip
\textbf{=== Stage 3: Second Round of Inference ===}

Use the fixed-position and category constraints:

- $b_1$ occurs at position $p_{[K]}$.

- $e_4=b_2$.

- $e_5\notin\{b_3,b_4\}$.

Combine these constraints with the previous stage and determine which assignments can be further restricted.

\medskip
\textbf{=== Stage 4: Final Inference ===}

a) Output the resulting tuples
[(Entity, Position, Attribute, Category), \ldots].

b) Verify that every constraint is satisfied.

c) Determine the attribute associated with category $b_3$.

\end{tcolorbox}

Under the local correspondence, entities correspond to people, ordered positions to floors, attributes to pets, and categories to occupations.
This intermediate representation is kept local and serves as the structural bridge for subsequent semantic rewriting and recovery.

\textbf{Semantic rewriting.}
\texttt{PriCon} then rewrites this constraint structure as a different story, obscuring the original task context while preserving the same reasoning relationships.
For example, the logic-grid task becomes a group-of-friends activity:

\begin{tcolorbox}[
    breakable,
    colback=myblue!5,
    colframe=myblue,
    boxrule=0.7pt,
    arc=3mm,
    left=4pt,
    right=4pt,
    top=3pt,
    bottom=3pt
]
\small\ttfamily
\linespread{1.03}\selectfont

Please complete the following four-part activity involving a group of friends.

\medskip
\textbf{=== Background ===}

There are [N] friends (Alice, Bob, Charlie, David, Emily), each living in a different neighborhood ordered by distance from the city center.
Each friend has a different favorite snack (chips/candy/popcorn/fruit/nuts) and a different hobby (painting/cycling/gaming/cooking/reading).

The following details are known:

1. Alice lives farther from the city center than Charlie.

2. The person who enjoys painting lives in neighborhood [P].

3. Bob's favorite snack is chips.

4. David's hobby is cooking.

5. The person who likes fruit lives farther from the city center than the person who likes popcorn.

6. Emily's hobby is neither cycling nor reading.

\medskip
\textbf{=== Part 1: Initial Gathering ===}

Gather all directly knowable details from the information provided without additional inference.

For example, ``Painter = neighborhood [P]'' and ``Bob's snack = chips'' are direct details, while ``Alice lives farther from the city center than Charlie'' is a relationship.

\medskip
\textbf{=== Part 2: First Round of Reasoning ===}

Use the relationship between Alice and Charlie and the relationship between fruit and popcorn:

- Given the ordered neighborhoods, which neighborhoods can Alice and Charlie each occupy?

- What does the ordering between fruit and popcorn imply?

- Combine these relations with Bob's preference for chips.

List the possible neighborhood ranges for the friends.

\medskip
\textbf{=== Part 3: Second Round of Reasoning ===}

Use the remaining hobby constraints:

- The person who enjoys painting lives in neighborhood [P].

- David enjoys cooking.

- Emily enjoys neither cycling nor reading.

Determine which possibilities are further restricted and which assignments can currently be established.

\medskip
\textbf{=== Part 4: Final Reasoning ===}

a) Output the resulting tuples
[(Name, Neighborhood, Snack, Hobby), \ldots].

b) Check whether every provided detail is satisfied.

c) What snack does the person who enjoys cycling prefer?

\end{tcolorbox}

The cloud therefore sees a group-of-friends reasoning task rather than the original logic-grid problem, while the local correspondence maps the returned reasoning back to the original entities and assignments.

\subsection{Operation Protection and Joint Protection}

Protecting the operation requires changing the apparent reasoning procedure while retaining a local path back to the original solution.

\textbf{Intermediate alternative formulation.}
To change the apparent operation, \texttt{PriCon} represents each feasible partial assignment as a state and each valid extension as a transition.
The original constraint-solving task is thereby converted into source-to-terminal path planning over a layered directed graph, with each path locally corresponding to a complete assignment.

\begin{tcolorbox}[
    breakable,
    colback=myorange!5,
    colframe=myorange,
    boxrule=0.7pt,
    arc=3mm,
    left=4pt,
    right=4pt,
    top=3pt,
    bottom=3pt
]
\small\ttfamily
\linespread{1.03}\selectfont

Please complete the following four-stage layered state-transition planning task.

\medskip
\textbf{=== Background ===}

A layered directed graph
$\mathcal{G}=({\mathcal{V}},{\mathcal{E}})$
contains layers
$\mathcal{L}_0,\mathcal{L}_1,\ldots,\mathcal{L}_{[N]}$.

The source node $v_0\in\mathcal{L}_0$ represents an empty state.
A node in layer $\mathcal{L}_k$ represents a feasible partial state after $k$ decisions.
A directed edge connects two states when the latter is a valid one-step extension of the former.

All exclusivity, ordering, fixed-position, and exclusion constraints have already been encoded into the admissible edges.
Nodes that violate any constraint are absent from the graph.

A set of valid terminal nodes
$\mathcal{V}_{T}\subseteq\mathcal{L}_{[N]}$
represents complete feasible states.

\medskip
\textbf{=== Stage 1: Initial Graph Inspection ===}

Identify the source node, terminal nodes, and admissible outgoing transitions from the first layer.

Record which states can be reached directly from the source.

\medskip
\textbf{=== Stage 2: Forward Reachability ===}

Propagate reachability layer by layer.

At each layer:

- retain only nodes reachable from the previous layer;

- remove transitions leading to dead-end states;

- record the remaining candidate paths.

\medskip
\textbf{=== Stage 3: Path Refinement ===}

Continue pruning states that cannot reach any valid terminal node.

Combine forward reachability with terminal reachability to identify transitions that can participate in at least one complete source-to-terminal path.

\medskip
\textbf{=== Stage 4: Path Recovery ===}

a) Return a complete valid path from $v_0$ to a node in $\mathcal{V}_{T}$.

b) Verify that every consecutive pair of states is connected by an admissible transition.

c) Return the transition labels along the selected path.

\end{tcolorbox}

This intermediate transformation changes the apparent reasoning procedure from logic-grid deduction to layered path planning while preserving a local mapping back to the original assignment.
The resulting path structure is then used for semantic rewriting.

\textbf{Semantic rewriting.}
\texttt{PriCon} then recasts the layered state-transition structure as a cooking-procedure planning task.
The cloud therefore reasons about feasible preparation steps rather than either logic-grid constraints or an explicit graph formulation.

\begin{tcolorbox}[
    breakable,
    colback=myorange!5,
    colframe=myorange,
    boxrule=0.7pt,
    arc=3mm,
    left=4pt,
    right=4pt,
    top=3pt,
    bottom=3pt
]
\small\ttfamily
\linespread{1.03}\selectfont

Please complete the following four-part cooking-procedure planning task.

\medskip
\textbf{=== Kitchen Setup ===}

You are preparing a multi-step tasting menu using a recipe-planning board with [N] preparation stages.

Each card on the board represents a valid partial kitchen state after some preparation steps have been completed.
An arrow between two cards represents an allowed next cooking action.
Ingredient conflicts, timing restrictions, incompatible techniques, and invalid preparation orders have already been removed from the board.

The process begins at the \emph{Kitchen Ready} card and must finish at one of the valid \emph{Meal Ready} cards.

\medskip
\textbf{=== Part 1: Inspect the Preparation Board ===}

Starting from \emph{Kitchen Ready}, identify all cooking actions that can be performed first.

Record the preparation states that can be reached after the first action.

\medskip
\textbf{=== Part 2: Explore Feasible Cooking Sequences ===}

Move through the preparation stages in order.

At each stage:

- keep only cooking states reachable from the previous stage;

- discard actions that lead to a dead end;

- retain preparation sequences that can still lead to a completed meal.

\medskip
\textbf{=== Part 3: Refine the Procedure ===}

Work backward from the valid \emph{Meal Ready} states and remove any preparation state that cannot lead to a completed menu.

Use both forward and backward feasibility to determine which cooking actions can belong to a valid procedure.

\medskip
\textbf{=== Part 4: Final Recipe Plan ===}

a) Return one complete cooking procedure from \emph{Kitchen Ready} to a valid \emph{Meal Ready} state.

b) Check that every cooking action follows an allowed transition on the preparation board.

c) List the action labels in the order in which they are performed.

\end{tcolorbox}

Here, preparation states and cooking actions correspond to states and transitions in the intermediate alternative formulation.
Locally, \texttt{PriCon} maps the returned cooking procedure back to the corresponding state-transition path and then to the original logic-grid solution.

\section{Experience Retrieval Details}
\label{app:experience-retrieval}

The local experience bank retrieves prior rewriting experiences through the intermediate mathematical representation.
We use this representation rather than the original or semantically rewritten task because it captures reusable solution structure while reducing dependence on surface-level domain descriptions.
\texttt{PriCon} standardizes it into four structural fields:
\emph{problem type}, \emph{decision variables}, \emph{objective}, and \emph{constraints}.
For a task $\mathcal{T}_i$, we denote this representation as
\[
Z_i =
\big(
z_i^{\mathrm{type}},
z_i^{\mathrm{var}},
z_i^{\mathrm{obj}},
z_i^{\mathrm{con}}
\big).
\]
For example, the medical-supply allocation task in Section~\ref{sec:intent-rewrite} can be represented as:
\begin{itemize}[leftmargin=1.5em, itemsep=0pt, topsep=2pt]
    \item \textbf{Problem type:} constrained resource allocation.
    \item \textbf{Decision variables:} discrete allocation amount for each demand-group/resource pair.
    \item \textbf{Objective:} minimize weighted shortage impact across demand groups.
    \item \textbf{Constraints:} allocation bounded by local demand; total allocation bounded by resource capacity; nonnegative discrete allocations.
\end{itemize}

This structured representation reduces sensitivity to surface-level differences in mathematical expressions.
For example,
$\min_x \sum_i w_i(d_i-x_i)^2$
and
$\min_x \sum_i w_i(x_i^2-2d_i x_i)$
have the same optimizer under the same constraints, since they differ only by the constant term $\sum_i w_i d_i^2$, although their raw expressions appear different.
Representing both formulations through the same four structural fields makes their underlying correspondence easier to capture.

\textbf{Field-wise similarity.}
For each field $k\in\{\mathrm{type},\mathrm{var},\mathrm{obj},\mathrm{con}\}$, we encode its text using an embedding model $E(\cdot)$ and compute cosine similarity:
\[
s_k(\mathcal{T}_i,\mathcal{T}_j)
=
\cos\!\left(
E(z_i^k),
E(z_j^k)
\right).
\]
The overall structural similarity between two tasks is the average over the four fields:
\[
S(\mathcal{T}_i,\mathcal{T}_j)
=
\frac{1}{4}
\sum_{k\in
\{\mathrm{type},\mathrm{var},\mathrm{obj},\mathrm{con}\}}
s_k(\mathcal{T}_i,\mathcal{T}_j).
\]
This field-wise comparison keeps retrieval simple and interpretable: two tasks receive a high score when their problem type, decision variables, objective, and constraints are jointly similar.

\textbf{Experience retrieval.}
For a new task $\mathcal{T}_q$, \texttt{PriCon} first derives its mathematical reformulation and corresponding four-field representation.
It then computes
\[
S_i = S(\mathcal{T}_q,\mathcal{T}_i)
\]
for each prior task $\mathcal{T}_i$ stored in the local experience bank $\mathcal{B}$ and ranks the corresponding experiences by $S_i$.
The top-$K$ candidates are retrieved as
\[
\mathcal{E}^{*}
=
\operatorname{TopK}_{\mathcal{E}_i\in\mathcal{B}}
\,S(\mathcal{T}_q,\mathcal{T}_i),
\]
When a sufficiently similar experience is available, its validated rewriting trajectory and correspondence $\mathcal{M}$ are reused to initialize the mathematical transformation and subsequent semantic rewriting for the new task.
Otherwise, \texttt{PriCon} constructs and validates a new surrogate from scratch and adds the resulting experience to the bank.
The four-field representation, embedding computation, retrieval, and experience storage are all performed locally.

\section{Task Screening Procedure}
\label{app:tasks}

We screen candidate tasks to focus on settings where cloud LLM assistance is genuinely needed.
Each candidate task is independently evaluated using GPT-4o-mini, DeepSeek-V4-Flash, and Qwen3.5-Flash without cloud consultation, \texttt{PriCon}, or the local experience bank.

A task is retained only if all three local SLMs fail to complete it end-to-end under the screening protocol.
We consider three major failure modes:
(i) \emph{context-length failure}, where the task exceeds the model's effective context capacity or necessary information is lost through truncation;
(ii) \emph{execution timeout}, where model inference or execution of the generated solution exceeds the prescribed time limit;
(iii) \emph{reasoning failure}, where the model produces a well-formed but incorrect output or is unable to complete the required reasoning steps; and
(iv) \emph{execution error}, where the generated solution fails during execution or does not produce a syntactically valid output.
The failure mode may differ across local SLMs for the same task.

Applying this screening criterion, we retain 10 instances from each task family, resulting in the 100-task benchmark used in our evaluation.

\section{category results}
\label{app:category-results}

{Appendix E reports per-family results for GPT-4o-mini (Table~\ref{tab:category-gpt4omini}), DeepSeek-V4-Flash (Table~\ref{tab:category-dsflash}), and Qwen3.5-Flash (Table~\ref{tab:category-qwen35flash}), complementing the aggregated results in Table~\ref{tab:main-results}. The trends observed in Table~\ref{tab:main-results}—near-perfect inference under direct consultation and sensitive-value removal, partial reduction under the decoy baseline, and near-zero inference under PriCon with semantic rewriting—hold consistently across individual task families.
}

\begin{table*}[h]
\centering
\captionsetup{justification=centering,font=scriptsize,skip=2pt}
\caption{Task-family-wise task-intent inference with \textbf{GPT-4o-mini} as the local SLM. Entries report Hit@1 / Hit@5 (\%).}
\label{tab:category-gpt4omini}
\tiny
\setlength{\aboverulesep}{0.5pt}
\setlength{\belowrulesep}{0.5pt}

\begin{tabular}{lc|cccccccc}
\toprule
\multirow{2}{*}{\textbf{Task Family}} &
\multirow{2}{*}{\makecell{\textbf{Inference}\\\textbf{Target}}} &
\multirow{2}{*}{\makecell{Direct cloud\\consultation}} &
\multirow{2}{*}{\makecell{Sensitive-value\\removal}} &
\multicolumn{2}{c}{Decoy} &
\multicolumn{2}{c}{\texttt{PriCon-Context}} &
\multicolumn{2}{c}{\texttt{PriCon-Joint}} \\
\cmidrule(lr){5-6}\cmidrule(lr){7-8}\cmidrule(lr){9-10}
& & & & 5 & 20 & Math-only & \textbf{Semantic} & Math-only & \textbf{Semantic} \\
\midrule
\multicolumn{10}{l}{\textbf{Structured and long-context reasoning}} \\
\midrule
\multirow{3}{*}{Alibi verification}
 & Context   & 100 / 100 & 92 / 92 & 24 / 33 & 1 / 1 & 0 / 0 & 0 / 0 & 0 / 0 & 0 / 0 \\
 & Operation & 100 / 100 & 100 / 100 & 49 / 63 & 25 / 56 & 69 / 77 & 61 / 66 & 0 / 0 & 0 / 0 \\
 & Joint     & 100 / 100 & 92 / 97 & 32 / 46 & 15 / 47 & 0 / 0 & 0 / 0 & 0 / 0 & 0 / 0 \\
\arrayrulecolor{gray!25}\hline\arrayrulecolor{black}
\multirow{3}{*}{Logic grid}
 & Context   & 97 / 100 & 100 / 100 & 19 / 24 & 1 / 9 & 0 / 0 & 0 / 0 & 16 / 36 & 0 / 0 \\
 & Operation & 97 / 100 & 100 / 100 & 31 / 65 & 28 / 66 & 83 / 88 & 61 / 66 & 5 / 11 & 0 / 0 \\
 & Joint     & 97 / 100 & 100 / 100 & 27 / 57 & 17 / 55 & 0 / 0 & 0 / 0 & 0 / 2 & 0 / 0 \\
\arrayrulecolor{gray!25}\hline\arrayrulecolor{black}
\multirow{3}{*}{Counterfactual reasoning}
 & Context   & 92 / 100 & 100 / 100 & 4 / 33 & 1 / 11 & 0 / 0 & 0 / 0 & 0 / 2 & 0 / 0 \\
 & Operation & 92 / 100 & 100 / 100 & 27 / 65 & 23 / 46 & 75 / 88 & 27 / 38 & 0 / 2 & 0 / 0 \\
 & Joint     & 92 / 100 & 100 / 100 & 19 / 50 & 14 / 38 & 0 / 0 & 0 / 0 & 0 / 0 & 0 / 0 \\
\arrayrulecolor{gray!25}\hline\arrayrulecolor{black}
\multirow{3}{*}{Recursive expansion}
 & Context   & 97 / 100 & 100 / 100 & 14 / 24 & 21 / 21 & 0 / 0 & 0 / 0 & 11 / 38 & 0 / 0 \\
 & Operation & 97 / 100 & 100 / 100 & 15 / 67 & 31 / 48 & 22 / 38 & 0 / 0 & 8 / 22 & 0 / 0 \\
 & Joint     & 97 / 97 & 100 / 100 & 0 / 51 & 18 / 40 & 0 / 0 & 0 / 0 & 0 / 8 & 0 / 0 \\
\arrayrulecolor{gray!25}\hline\arrayrulecolor{black}
\multirow{3}{*}{Relation graph}
 & Context   & 92 / 97 & 100 / 100 & 4 / 39 & 6 / 21 & 0 / 0 & 0 / 0 & 2 / 13 & 0 / 0 \\
 & Operation & 100 / 100 & 100 / 100 & 39 / 67 & 10 / 53 & 33 / 55 & 66 / 86 & 2 / 8 & 0 / 0 \\
 & Joint     & 92 / 97 & 100 / 100 & 25 / 58 & 6 / 45 & 0 / 0 & 0 / 0 & 2 / 2 & 0 / 0 \\
\arrayrulecolor{gray!25}\hline\arrayrulecolor{black}
\midrule
\multicolumn{10}{l}{\textbf{Decision and optimization}} \\
\midrule
\multirow{3}{*}{Financial modeling}
 & Context   & 97 / 100 & 100 / 100 & 13 / 24 & 1 / 4 & 11 / 11 & 0 / 0 & 0 / 0 & 0 / 0 \\
 & Operation & 100 / 100 & 100 / 100 & 47 / 67 & 17 / 66 & 58 / 69 & 16 / 22 & 0 / 2 & 0 / 0 \\
 & Joint     & 97 / 100 & 100 / 100 & 43 / 57 & 10 / 55 & 0 / 2 & 0 / 0 & 0 / 0 & 0 / 0 \\
\arrayrulecolor{gray!25}\hline\arrayrulecolor{black}
\multirow{3}{*}{Budget allocation}
 & Context   & 95 / 100 & 82 / 100 & 14 / 45 & 18 / 68 & 0 / 8 & 0 / 0 & 0 / 2 & 0 / 0 \\
 & Operation & 100 / 100 & 97 / 100 & 29 / 60 & 13 / 33 & 72 / 80 & 33 / 50 & 0 / 2 & 0 / 0 \\
 & Joint     & 97 / 100 & 85 / 100 & 25 / 50 & 8 / 28 & 0 / 0 & 0 / 0 & 0 / 0 & 0 / 0 \\
\arrayrulecolor{gray!25}\hline\arrayrulecolor{black}
\multirow{3}{*}{Office setup}
 & Context   & 90 / 97 & 82 / 90 & 0 / 48 & 1 / 28 & 2 / 13 & 0 / 0 & 0 / 0 & 0 / 0 \\
 & Operation & 97 / 97 & 92 / 97 & 0 / 65 & 4 / 51 & 19 / 36 & 5 / 30 & 0 / 0 & 0 / 0 \\
 & Joint     & 92 / 97 & 82 / 92 & 0 / 54 & 2 / 43 & 0 / 0 & 0 / 0 & 0 / 0 & 0 / 0 \\
\arrayrulecolor{gray!25}\hline\arrayrulecolor{black}
\midrule
\multicolumn{10}{l}{\textbf{Coding}} \\
\midrule
\multirow{3}{*}{Python coding}
 & Context   & 97 / 100 & 95 / 100 & 0 / 9 & 1 / 11 & 0 / 0 & 0 / 0 & 0 / 12 & 0 / 0 \\
 & Operation & 100 / 100 & 100 / 100 & 23 / 65 & 1 / 21 & 32 / 50 & 7 / 12 & 0 / 5 & 0 / 0 \\
 & Joint     & 97 / 100 & 97 / 100 & 19 / 51 & 1 / 17 & 0 / 0 & 0 / 0 & 0 / 2 & 0 / 0 \\
\arrayrulecolor{gray!25}\hline\arrayrulecolor{black}
\midrule
\multicolumn{10}{l}{\textbf{Cross-domain composite}} \\
\midrule
\multirow{3}{*}{Composite cross-domain}
 & Context   & 77 / 87 & 90 / 92 & 2 / 37 & 6 / 31 & 0 / 0 & 0 / 0 & 0 / 2 & 0 / 0 \\
 & Operation & 90 / 100 & 97 / 100 & 35 / 63 & 23 / 56 & 27 / 44 & 36 / 50 & 0 / 0 & 0 / 0 \\
 & Joint     & 82 / 92 & 85 / 90 & 16 / 40 & 14 / 47 & 0 / 0 & 0 / 0 & 0 / 0 & 0 / 0 \\
\arrayrulecolor{gray!25}\hline\arrayrulecolor{black}
\bottomrule
\end{tabular}
\end{table*}

\begin{table*}[h]
\centering
\captionsetup{justification=centering,font=scriptsize,skip=2pt}
\caption{Task-family-wise task-intent inference with \textbf{DeepSeek-V4-Flash} as the local SLM. Entries report Hit@1 / Hit@5 (\%).}
\label{tab:category-dsflash}
\tiny
\setlength{\aboverulesep}{0.5pt}
\setlength{\belowrulesep}{0.5pt}

\begin{tabular}{lc|cccccccc}
\toprule
\multirow{2}{*}{\textbf{Task Family}} &
\multirow{2}{*}{\makecell{\textbf{Inference}\\\textbf{Target}}} &
\multirow{2}{*}{\makecell{Direct cloud\\consultation}} &
\multirow{2}{*}{\makecell{Sensitive-value\\removal}} &
\multicolumn{2}{c}{Decoy} &
\multicolumn{2}{c}{\texttt{PriCon-Context}} &
\multicolumn{2}{c}{\texttt{PriCon-Joint}} \\
\cmidrule(lr){5-6}\cmidrule(lr){7-8}\cmidrule(lr){9-10}
& & & & 5 & 20 & Math-only & \textbf{Semantic} & Math-only & \textbf{Semantic} \\
\midrule
\multicolumn{10}{l}{\textbf{Structured and long-context reasoning}} \\
\midrule
\multirow{3}{*}{Alibi verification}
 & Context   & 97 / 100 & 90 / 100 & 29 / 46 & 1 / 0 & 0 / 0 & 0 / 0 & 0 / 0 & 0 / 0 \\
 & Operation & 100 / 100 & 100 / 100 & 46 / 64 & 44 / 70 & 47 / 66 & 63 / 94 & 0 / 2 & 0 / 0 \\
 & Joint     & 95 / 100 & 87 / 100 & 42 / 53 & 20 / 39 & 0 / 0 & 0 / 0 & 0 / 0 & 0 / 0 \\
\arrayrulecolor{gray!25}\hline\arrayrulecolor{black}
\multirow{3}{*}{Logic grid}
 & Context   & 97 / 100 & 100 / 100 & 5 / 17 & 4 / 15 & 0 / 0 & 0 / 0 & 8 / 22 & 0 / 0 \\
 & Operation & 92 / 97 & 97 / 100 & 38 / 66 & 47 / 85 & 61 / 80 & 58 / 72 & 2 / 11 & 0 / 0 \\
 & Joint     & 100 / 100 & 100 / 100 & 25 / 61 & 22 / 47 & 0 / 0 & 0 / 0 & 0 / 8 & 0 / 0 \\
\arrayrulecolor{gray!25}\hline\arrayrulecolor{black}
\multirow{3}{*}{Counterfactual reasoning}
 & Context   & 100 / 100 & 100 / 100 & 0 / 28 & 1 / 0 & 0 / 0 & 0 / 0 & 0 / 0 & 0 / 0 \\
 & Operation & 100 / 100 & 100 / 100 & 10 / 66 & 32 / 68 & 77 / 88 & 94 / 100 & 5 / 8 & 0 / 0 \\
 & Joint     & 100 / 100 & 100 / 100 & 9 / 49 & 15 / 38 & 0 / 0 & 0 / 0 & 0 / 0 & 0 / 0 \\
\arrayrulecolor{gray!25}\hline\arrayrulecolor{black}
\multirow{3}{*}{Recursive expansion}
 & Context   & 97 / 100 & 100 / 100 & 5 / 27 & 1 / 5 & 0 / 0 & 0 / 0 & 0 / 16 & 0 / 0 \\
 & Operation & 85 / 97 & 92 / 97 & 8 / 66 & 16 / 53 & 25 / 44 & 72 / 86 & 2 / 13 & 0 / 0 \\
 & Joint     & 97 / 100 & 100 / 100 & 0 / 61 & 7 / 29 & 0 / 0 & 0 / 0 & 0 / 13 & 0 / 0 \\
\arrayrulecolor{gray!25}\hline\arrayrulecolor{black}
\multirow{3}{*}{Relation graph}
 & Context   & 100 / 100 & 100 / 100 & 0 / 30 & 4 / 15 & 0 / 0 & 0 / 0 & 0 / 0 & 0 / 0 \\
 & Operation & 100 / 100 & 100 / 100 & 38 / 66 & 36 / 80 & 63 / 75 & 72 / 91 & 2 / 2 & 0 / 0 \\
 & Joint     & 100 / 100 & 100 / 100 & 22 / 61 & 17 / 45 & 0 / 0 & 0 / 0 & 0 / 0 & 0 / 0 \\
\arrayrulecolor{gray!25}\hline\arrayrulecolor{black}
\midrule
\multicolumn{10}{l}{\textbf{Decision and optimization}} \\
\midrule
\multirow{3}{*}{Financial modeling}
 & Context   & 100 / 100 & 100 / 100 & 23 / 40 & 4 / 5 & 11 / 11 & 0 / 0 & 0 / 0 & 0 / 0 \\
 & Operation & 100 / 100 & 100 / 100 & 43 / 61 & 26 / 73 & 61 / 66 & 72 / 75 & 0 / 0 & 0 / 0 \\
 & Joint     & 100 / 100 & 100 / 100 & 38 / 55 & 12 / 40 & 0 / 13 & 0 / 0 & 0 / 0 & 0 / 0 \\
\arrayrulecolor{gray!25}\hline\arrayrulecolor{black}
\multirow{3}{*}{Budget allocation}
 & Context   & 97 / 100 & 92 / 100 & 13 / 43 & 23 / 65 & 0 / 0 & 0 / 0 & 2 / 11 & 0 / 0 \\
 & Operation & 97 / 100 & 92 / 100 & 21 / 64 & 26 / 68 & 61 / 72 & 75 / 86 & 0 / 2 & 0 / 0 \\
 & Joint     & 97 / 100 & 92 / 100 & 17 / 52 & 12 / 38 & 0 / 0 & 0 / 0 & 0 / 2 & 0 / 0 \\
\arrayrulecolor{gray!25}\hline\arrayrulecolor{black}
\multirow{3}{*}{Office setup}
 & Context   & 92 / 100 & 90 / 95 & 0 / 57 & 21 / 50 & 16 / 16 & 0 / 0 & 0 / 0 & 0 / 0 \\
 & Operation & 87 / 95 & 80 / 90 & 0 / 64 & 12 / 55 & 19 / 36 & 22 / 47 & 2 / 2 & 0 / 0 \\
 & Joint     & 92 / 100 & 90 / 95 & 0 / 58 & 5 / 31 & 0 / 0 & 0 / 0 & 0 / 0 & 0 / 0 \\
\arrayrulecolor{gray!25}\hline\arrayrulecolor{black}
\midrule
\multicolumn{10}{l}{\textbf{Coding}} \\
\midrule
\multirow{3}{*}{Python coding}
 & Context   & 100 / 100 & 100 / 100 & 0 / 0 & 11 / 10 & 0 / 0 & 0 / 0 & 0 / 0 & 0 / 0 \\
 & Operation & 100 / 100 & 100 / 100 & 26 / 58 & 14 / 48 & 30 / 37 & 25 / 47 & 2 / 5 & 0 / 0 \\
 & Joint     & 100 / 100 & 97 / 100 & 19 / 46 & 6 / 27 & 0 / 0 & 0 / 0 & 0 / 0 & 0 / 0 \\
\arrayrulecolor{gray!25}\hline\arrayrulecolor{black}
\midrule
\multicolumn{10}{l}{\textbf{Cross-domain composite}} \\
\midrule
\multirow{3}{*}{Composite cross-domain}
 & Context   & 95 / 100 & 87 / 97 & 11 / 47 & 6 / 30 & 0 / 0 & 0 / 0 & 0 / 0 & 0 / 0 \\
 & Operation & 90 / 100 & 90 / 100 & 25 / 49 & 41 / 75 & 36 / 52 & 36 / 50 & 0 / 0 & 0 / 0 \\
 & Joint     & 97 / 100 & 87 / 97 & 17 / 45 & 19 / 42 & 0 / 0 & 0 / 0 & 0 / 0 & 0 / 0 \\
\arrayrulecolor{gray!25}\hline\arrayrulecolor{black}
\bottomrule
\end{tabular}
\end{table*}

\begin{table*}[h]
\centering
\captionsetup{justification=centering,font=scriptsize,skip=2pt}
\caption{Task-family-wise task-intent inference with \textbf{Qwen3.5-Flash} as the local SLM. Entries report Hit@1 / Hit@5 (\%).}
\label{tab:category-qwen35flash}
\tiny
\setlength{\aboverulesep}{0.5pt}
\setlength{\belowrulesep}{0.5pt}

\begin{tabular}{lc|cccccccc}
\toprule
\multirow{2}{*}{\textbf{Task Family}} &
\multirow{2}{*}{\makecell{\textbf{Inference}\\\textbf{Target}}} &
\multirow{2}{*}{\makecell{Direct cloud\\consultation}} &
\multirow{2}{*}{\makecell{Sensitive-value\\removal}} &
\multicolumn{2}{c}{Decoy} &
\multicolumn{2}{c}{\texttt{PriCon-Context}} &
\multicolumn{2}{c}{\texttt{PriCon-Joint}} \\
\cmidrule(lr){5-6}\cmidrule(lr){7-8}\cmidrule(lr){9-10}
& & & & 5 & 20 & Math-only & \textbf{Semantic} & Math-only & \textbf{Semantic} \\
\midrule
\multicolumn{10}{l}{\textbf{Structured and long-context reasoning}} \\
\midrule
\multirow{3}{*}{Alibi verification}
 & Context   & 100 / 100 & 95 / 100 & 10 / 49 & 0 / 0 & 0 / 0 & 0 / 0 & 0 / 0 & 0 / 0 \\
 & Operation & 100 / 100 & 100 / 100 & 54 / 67 & 39 / 61 & 41 / 61 & 75 / 86 & 0 / 8 & 0 / 2 \\
 & Joint     & 100 / 100 & 100 / 100 & 36 / 52 & 32 / 55 & 0 / 0 & 0 / 0 & 0 / 0 & 0 / 0 \\
\arrayrulecolor{gray!25}\hline\arrayrulecolor{black}
\multirow{3}{*}{Logic grid}
 & Context   & 100 / 100 & 97 / 100 & 0 / 38 & 0 / 0 & 0 / 0 & 0 / 0 & 11 / 19 & 0 / 0 \\
 & Operation & 100 / 100 & 97 / 100 & 19 / 67 & 26 / 56 & 63 / 88 & 47 / 63 & 5 / 8 & 0 / 0 \\
 & Joint     & 100 / 100 & 97 / 100 & 8 / 61 & 22 / 50 & 0 / 0 & 0 / 0 & 0 / 0 & 0 / 0 \\
\arrayrulecolor{gray!25}\hline\arrayrulecolor{black}
\multirow{3}{*}{Counterfactual reasoning}
 & Context   & 95 / 100 & 100 / 100 & 0 / 36 & 0 / 1 & 0 / 0 & 0 / 0 & 0 / 0 & 0 / 0 \\
 & Operation & 100 / 100 & 97 / 100 & 13 / 67 & 19 / 44 & 75 / 91 & 86 / 100 & 2 / 11 & 0 / 0 \\
 & Joint     & 100 / 100 & 100 / 100 & 4 / 52 & 16 / 39 & 0 / 0 & 0 / 0 & 0 / 2 & 0 / 0 \\
\arrayrulecolor{gray!25}\hline\arrayrulecolor{black}
\multirow{3}{*}{Recursive expansion}
 & Context   & 100 / 100 & 97 / 100 & 10 / 12 & 0 / 0 & 0 / 0 & 0 / 0 & 11 / 27 & 0 / 0 \\
 & Operation & 100 / 100 & 100 / 100 & 30 / 67 & 4 / 41 & 8 / 36 & 50 / 55 & 13 / 22 & 0 / 0 \\
 & Joint     & 100 / 100 & 97 / 100 & 25 / 61 & 4 / 37 & 0 / 0 & 0 / 0 & 2 / 13 & 0 / 0 \\
\arrayrulecolor{gray!25}\hline\arrayrulecolor{black}
\multirow{3}{*}{Relation graph}
 & Context   & 100 / 100 & 95 / 97 & 10 / 51 & 9 / 26 & 0 / 0 & 0 / 0 & 0 / 0 & 0 / 0 \\
 & Operation & 100 / 100 & 85 / 85 & 32 / 67 & 14 / 41 & 61 / 72 & 75 / 88 & 0 / 5 & 0 / 0 \\
 & Joint     & 100 / 100 & 97 / 97 & 8 / 61 & 12 / 37 & 0 / 0 & 0 / 0 & 0 / 0 & 0 / 0 \\
\arrayrulecolor{gray!25}\hline\arrayrulecolor{black}
\midrule
\multicolumn{10}{l}{\textbf{Decision and optimization}} \\
\midrule
\multirow{3}{*}{Financial modeling}
 & Context   & 97 / 100 & 100 / 100 & 0 / 12 & 0 / 3 & 11 / 11 & 0 / 0 & 0 / 0 & 0 / 0 \\
 & Operation & 100 / 100 & 100 / 100 & 58 / 66 & 17 / 46 & 66 / 77 & 58 / 69 & 0 / 5 & 0 / 0 \\
 & Joint     & 97 / 100 & 100 / 100 & 57 / 60 & 14 / 41 & 0 / 0 & 0 / 0 & 0 / 0 & 0 / 0 \\
\arrayrulecolor{gray!25}\hline\arrayrulecolor{black}
\multirow{3}{*}{Budget allocation}
 & Context   & 95 / 97 & 77 / 100 & 10 / 56 & 15 / 67 & 0 / 0 & 0 / 0 & 2 / 2 & 0 / 0 \\
 & Operation & 100 / 100 & 95 / 100 & 30 / 62 & 17 / 44 & 66 / 72 & 61 / 80 & 0 / 2 & 0 / 0 \\
 & Joint     & 97 / 97 & 80 / 97 & 21 / 50 & 14 / 39 & 0 / 0 & 0 / 0 & 0 / 0 & 0 / 0 \\
\arrayrulecolor{gray!25}\hline\arrayrulecolor{black}
\multirow{3}{*}{Office setup}
 & Context   & 87 / 90 & 82 / 95 & 10 / 64 & 9 / 47 & 2 / 19 & 0 / 0 & 0 / 11 & 0 / 0 \\
 & Operation & 95 / 100 & 92 / 95 & 6 / 64 & 7 / 41 & 25 / 38 & 47 / 58 & 0 / 0 & 2 / 5 \\
 & Joint     & 90 / 92 & 85 / 97 & 6 / 56 & 6 / 37 & 0 / 0 & 0 / 0 & 0 / 0 & 0 / 2 \\
\arrayrulecolor{gray!25}\hline\arrayrulecolor{black}
\midrule
\multicolumn{10}{l}{\textbf{Coding}} \\
\midrule
\multirow{3}{*}{Python coding}
 & Context   & 95 / 100 & 95 / 100 & 10 / 12 & 9 / 18 & 0 / 0 & 0 / 0 & 2 / 12 & 0 / 0 \\
 & Operation & 100 / 100 & 100 / 100 & 19 / 67 & 7 / 19 & 30 / 47 & 30 / 42 & 0 / 7 & 0 / 0 \\
 & Joint     & 95 / 100 & 95 / 100 & 17 / 56 & 6 / 17 & 0 / 0 & 0 / 0 & 0 / 2 & 0 / 0 \\
\arrayrulecolor{gray!25}\hline\arrayrulecolor{black}
\midrule
\multicolumn{10}{l}{\textbf{Cross-domain composite}} \\
\midrule
\multirow{3}{*}{Composite cross-domain}
 & Context   & 77 / 82 & 87 / 95 & 22 / 44 & 1 / 17 & 0 / 0 & 0 / 0 & 0 / 0 & 0 / 0 \\
 & Operation & 97 / 100 & 92 / 100 & 23 / 59 & 27 / 61 & 38 / 55 & 33 / 50 & 0 / 2 & 0 / 0 \\
 & Joint     & 80 / 82 & 80 / 95 & 17 / 43 & 22 / 55 & 0 / 0 & 0 / 0 & 0 / 0 & 0 / 0 \\
\arrayrulecolor{gray!25}\hline\arrayrulecolor{black}
\bottomrule
\end{tabular}
\end{table*}

\end{document}